\documentclass[10pt]{article}

\usepackage{amsmath}
\usepackage{amssymb}
\usepackage{amsfonts,bbold} 

\usepackage{graphicx}

\usepackage{natbib} 
\usepackage{color} 

\usepackage[toc,page]{appendix} 

\usepackage[labelfont=bf,labelsep=period]{caption} 

\makeatletter
\renewcommand{\@biblabel}[1]{\quad#1.}
\makeatother

\date{}

\newcommand{\ba}{\mathbf{a}}
\newcommand{\bs}{\mathbf{s}}
\newcommand{\Ntot}{N_\mathrm{tot}}
\newcommand{\fhat}{\widehat{f}}
\newcommand{\fstar}{f^\star}
\newcommand{\Zstar}{{Z^\star}}

\newcommand{\nN}{N}
\newcommand{\nK}{K}

\newcommand{\wstar}{w^\star}

\newcommand{\pistar}{\pi^\star}

\newcommand{\bC}{\mathbf{C}}
\newcommand{\bbeta}{\boldsymbol{\beta}}
\newcommand{\bb}{\boldsymbol{b}}
\newcommand{\bGam}{\boldsymbol{\Gamma}}
\newcommand{\bgam}{\boldsymbol{\gamma}}
\newcommand{\bpi}{\boldsymbol{\pi}}
\newcommand{\bpistar}{\boldsymbol{\pi}^\star}
\newcommand{\tR}{t_R} 
\newcommand{\tRstar}{{t_R^\star}} 
\newcommand{\Kapstar}{{\mathcal{K}^\star}}
\newcommand{\Kap}{{\mathcal{K}}}

\newcommand{\Wstar}{W^\star}

\newcommand{\PSNR}{\mathrm{P}_Z}
\newcommand{\PPCV}{\mathrm{P}_W}
\newcommand{\aSNR}{\alpha_Z}
\newcommand{\aPCV}{\alpha_W}

\newcommand{\Kest}{\breve{K}}
\newcommand{\West}{\breve{W}}
\newcommand{\Wbar}{\overline{W}}

\newcommand{\pP}{\mathrm{P}} 
\newcommand{\pE}{\mathrm{E}} 
\newcommand{\Var}{\mathrm{Var}}
\newcommand{\Cov}{\mathrm{Cov}}

\newcommand{\om}{\omega}
\newcommand{\Ome}{\mathbb{\Omega}}
\newcommand{\Id}{\mathrm{\bf Id}}
\newcommand{\bu}{\mathbf{u}}

\newcommand{\eE}{\mathcal{E}}
\newcommand{\bPhi}{\boldsymbol{\Phi}}
\newcommand{\boeta}{\boldsymbol{\eta}}
\newcommand{\bLam}{\boldsymbol{\Lambda}}
\newcommand{\bU}{\boldsymbol{\rm U}}
\newcommand{\bV}{\boldsymbol{\rm V}}
\newcommand{\bA}{\boldsymbol{\rm A}}
\newcommand{\bzeta}{\boldsymbol{\zeta}}
\newcommand{\bD}{\boldsymbol{\rm D}}
\newcommand{\bX}{\boldsymbol{\rm X}}
\newcommand{\bT}{\boldsymbol{\rm T}}
\newcommand{\bP}{\boldsymbol{\rm P}}
\newcommand{\bDelta}{\boldsymbol{\Delta}}
\newcommand{\bd}{\mathbf{d}}

\newcommand{\refeq}[1]{{eq.~\ref{eq:#1}}}
\newcommand{\refsec}[1]{{section~\ref{sec:#1}}}
\newcommand{\reffig}[1]{{Fig.~\ref{fig:#1}}}

\title{Percept formation from neural populations in sensory decision-making tasks}
\author{Adrien Wohrer$^{1,\ast}$, Christian Machens$^{2}$}

\begin{document}


\maketitle 
\begin{flushleft}

\bf{1} Group for Neural Theory, INSERM U960, \'Ecole Normale Supérieure, Paris, France
\\
\bf{2} Champalimaud Neuroscience Program, Libson, Portugal
\\
$\ast$ E-mail: adrien.wohrer@ens.fr
\end{flushleft}


\begin{abstract}
We study a standard linear readout model of perceptual integration from a population of sensory neurons. We show that the readout can be associated to a set of characteristic equations which summarize the joint trial-to-trial covariance structure of neural activities and animal percept. These characteristic equations implicitly determine the readout parameters that were used by the animal to create its percept. In particular, they implicitly constrain the temporal integration window $w$ and the typical number of neurons $K$ which give rise to the percept. Comparing neural and behavioral sensitivity alone cannot disentangle these two sources of perceptual integration, so the characteristic equations also involve a measure of choice signals, like those assessed by the classic experimental measure of choice probabilities. We then propose a statistical method of analysis which allows to recover the typical scales of integration $w$ and $K$ from finite numbers of recorded neurons and recording trials, and show the efficiency of this method on an artificial encoding network. We also study the statistical method theoretically, and relate its laws of convergence to the underlying structure of neural activity in the population, as described through its singular value decomposition. Altogether, our method provides the first thorough interpretation of feedforward percept formation from a population of sensory neurons. It can readily be applied to experimental recordings in classic sensory decision-making tasks, and hopefully provide new insights into the nature of perceptual integration.
\end{abstract}

%
%

\section{Introduction}

Most cortical neurons are noisy, or at least appear so to experimenters. When a sensory neuron's spikes are recorded in response to a well-controlled stimulus, they will show a large variability from trial to trial. This noisiness has been acknowledged from early on, as a nuisance preventing experimenters from easy access to the encoding properties of sensory neurons. 
But what is the impact of trial-to-trial sensory noise on the organism itself? This question gained renewed interest a few decades ago, with the generalization of experimental setups recording neural activity from awake, behaving animals \citep{Mountcastle1990,Britten1992}. In these setups, animals are presented with a set of stimuli $f$ and trained to respond differentially to different values of $f$, thus providing an (indirect) report of their percept of $f$. As neural activity and animal behavior are simultaneously monitored, it becomes possible to seek a causal link between the two.

In such setups, one particular hypothesis---which we refer to as the ``sensory noise'' hypothesis---has proven instrumental in linking neural activity and percepts. It postulates that trial-to-trial noise at the level of sensory neurons is the main factor limiting the accuracy of the animal's perceptual judgements \citep{Werner1965,Talbot1968}. Indeed, signal detection theory provides the adequate tools to estimate such accuracies. Any type of biological response to a stimulus $f$---say $r(f)$---can be associated to a signal-to-noise ratio (SNR), which measures how typical variations in $r$ due to a change of stimulus $f$ (the {\it signal}) compare to intrinsic variations of $r$ from trial to trial (the {\it noise}). When $r(f)$ measures the response of a neuron to stimulus $f$, the resulting SNR is often called the {\it neurometric} sensitivity for that particular neuron. Alternatively, $r(f)$ may also be the response of the animal itself to stimulus $f$. The resulting SNR is called the animal's {\it psychometric} sensitivity, which quantifies the animal's ability to discriminate nearby stimulus values $f$. Reformulated in terms of SNRs, the ``sensory noise'' hypothesis states that neurometric sensitivity, computed from the population of sensory neurons under survey, is equal to the psychometric sensitivity for the animal in the task.

Applying this idea, neurometric and psychometric sensitivities have often been computed and compared, in various sensory systems and behavioral tasks \citep[see, e.g.,][for reference]{Romo2003,Gold2007}.
However, it was progressively realized that most of these comparisons bear no simple interpretation, because the neurometric sensitivity is not a fixed quantity: it depends on how information is read out from the neurons. 
For example, if the various sensory neurons in the population behave independently one from another, then the overall SNR from the population will essentially be the sum of individual SNRs and thus, the experimenter's estimate of neurometric sensitivity will depend on how many neurons---say $\nK$---they included in their analysis. This intuition still holds in realistic populations where neurons are not independent, with the additional complexity that the evolution of neural SNR with $\nK$ is very influenced by the correlation structure of noise in the population \citep{Shadlen1998, Abbott1999, Averbeck2006}.

More subtly, another parameter has a direct influence on estimated neurometric SNRs: the time scale $w$ used to integrate each neuron's spike train, to describe the neuron's activity over the trial \citep{Cohen2009}. Indeed, through the central limit theorem, the more neural spikes are integrated into the readout, the more accurate that readout will be. Adding extra neurons through $\nK$, or extra spikes for each neuron through $w$, will thus have the same type of impact on the readout's overall SNR. In fact, if all neurons from the population are identical, independent Poisson encoders, one can easily show that the readout's overall SNR scales with $\sqrt{w\nK}$, emphasizing the duality between $\nK$ and $w$.

As there is no unique way of reading out information from a population of sensory neurons, a question naturally arises: what type of readout does the organism use? For example, how many sensory neurons $\nK$, and what typical integration time scale $w$, provide a relevant description of the animal's percept formation? The ``sensory noise'' hypothesis can precisely be used to answer this question: the `true' neuronal readout for the organism must be the one providing the best account of animal behavior. However, the previous $\nK$--$w$ discussion clearly shows that comparing neurometric SNR to psychometric SNR is not sufficient to target the true readout: there will be several combinations of $\nK$ and $w$ leading to the same overall neurometric SNR, while corresponding to very different extraction strategies by the animal. Thus, an additional experimental measure is required to recover the typical scales of integration of the true readout.

{\it Choice signals} are a good candidate for this additional measure. 
In two-alternative tasks, where the animal must report a binary discrimination of stimulus value (say, $f>0$ or $f<0$), choice signals are generally computed in the form of {\it choice probabilities} (CP) \citep{Green1966,Britten1996}. CP is computed for each recorded neuron individually, and quantifies the trial-to-trial correlation between the activity of that neuron and the animal's ultimate (binary) choice on the trial, all other features being held constant. In particular, since CP is computed across trials with the same stimulus value (generally uninformative, i.e., $f=0$), the observed correlations cannot reflect the influence of stimulus on neural activity and animal choice. Instead, a significant CP can only result from the process by which the neuron's activity influences---or is influenced by---the animal's forming perceptual decision. 

It is intuitively clear that CPs reveal something about the way information is extracted from sensory neurons. For example, if the animal's percept is built from a single neuron, then that neuron will have a very large CP, because its activity on every trial directly predicts the animal's percept. Instead, if several independent neurons contribute to form the animal's percept, then they are all expected to have low CP value, as the activity of each neuron has only a marginal impact on the animal's decision. However, converting this intuition about choice signals into a quantitative interpretation was long hampered by the fact that, just like neurometric SNR, CP values are largely influenced by the population's noise covariance structure. For example, a neuron may not be utilized by the animal to form its percept, and yet display significant CP because its activity is correlated with that of another neuron being utilized. As a result, early studies relating CP values to the animal's perceptual readout only relied on numerical simulations \citep{Shadlen1996,Cohen2009}, assuming very specific noise correlation structures that weakened the generalizations of their results.
Only very recently have \citet{Haefner2013} provided an analytical expression for CP values in the presence of noise correlations (see \refsec{CP}), opening the door to general, quantitative interpretations of choice probabilities.

In this article, we show how the combined information of animal sensitivity (SNR) and choice signals allows to estimate the typical scales of percept formation by the animal, both across neurons (number of neurons involved $\nK$) and in time (integration window $w$). Our results apply in the standard feedforward model of percept formation, and can be derived for any noise covariance structure in the neural population. We first show how the joint covariance structure of neural activities and animal percept leads to a set of characteristic equations for the readout, which implicitly determine the animal's perceptual readout policy across neurons and time. Then, we show how these characteristic equations can be used in a statistical form, across the ensemble of trials and neurons available to the experimenter, to determine the typical scales $\nK$ and $w$ of percept formation from the activity of sensory neurons. This approach is mandatory since experimental measurements can only provide statistical samples of the full neural population.
Using an artificial neural network to provide sensory encoding, we show that our method can reliably recover the true scales of perceptual integration, without requiring full measurement of the neural population. Thus, our method can readily be applied to real experimental data, and provide new insights into the nature of sensory percept formation.


%
%

\section{Methods}

\subsection{Framework and notations}

\begin{figure}
\centerline{
         \includegraphics[width=.9\columnwidth]{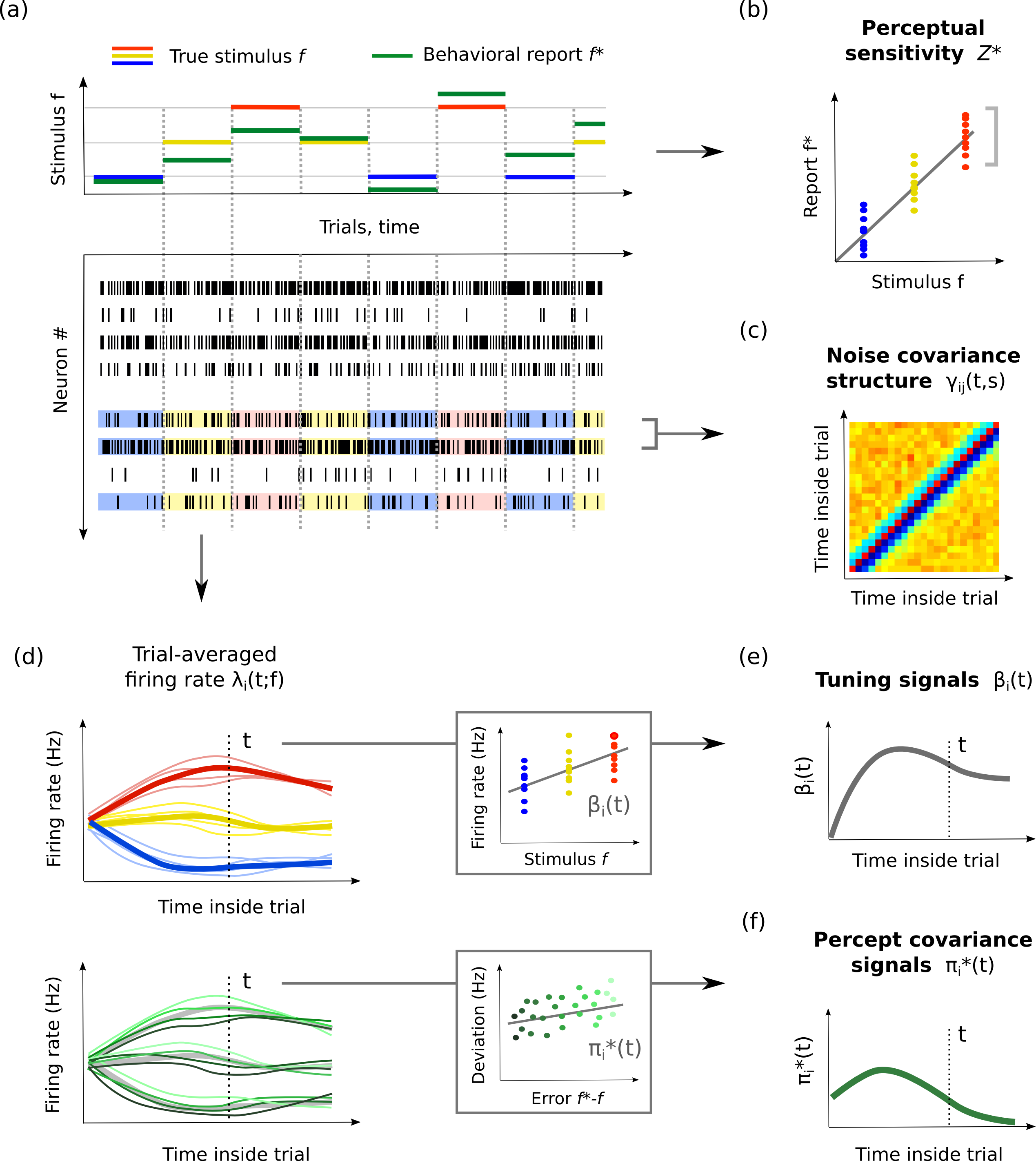}
	}
        \caption{\label{fig:frame} Framework and main experimental measures. (a) Experimental setup. Top: A set of stimulus values $f$ (color-coded as blue, yellow, red) are repeatedly presented to an animal, which reports its percept $\fstar$ on each trial (color-coded as green). Bottom: In each session, several task-relevant sensory neurons are recorded simultaneously with behavior. (b) Perceptual sensitivity $\Zstar$ is defined as the square SNR of the animal's reports $\fstar(f)$. (c) The noise covariance structure can be assessed in each each pair of simultaneously recorded neurons, as their joint peri-stimulus histogram (JPSTH). (d) Trial-wise response of a particular neuron. Each thin line is the schematical representation of the spike train on each trial. Segregating trials according to stimulus (top), we access the neuron's peri-stimulus histogram (PSTH) and its tuning curve---shown in panel (e). Segregating trials according to the animal's perceptual error $\fstar(f)-f$ (bottom), we access the neuron's percept covariance (PCV) curve---shown in panel (f).}
\end{figure}

We place ourselves in a general framework, describing a typical perceptual decision-making experiment (\reffig{frame}). On each trial, a different stimulus $f$ is presented to the animal (\reffig{frame}a, top), which then takes a decision according to its internal judgement $\fstar$ of stimulus value. Our framework assumes that this percept $\fstar$ is directly available to the experimenter on each trial. In real experimental setups, the animal's report is generally more indirect---typically a binary choice based on the unknown percept $\fstar$. We choose the former approach because it applies generically to most perceptual decision-making experiments, whereas the ``choice'' part is more dependent on each particular setup. We detail later how both approaches can be reconciled through simple models of the animal's behavior (\refsec{CP}).

Simultaneously, experimenters record neural activities from a large population of sensory neurons, which is assumed to convey the basic information about $f$ used by the animal to take its decision (\reffig{frame}a, bottom). Typical examples could be area MT in the context of a moving dot discrimination task \citep[e.g.,][]{Britten1992}, area MT or V2 in the context of a depth discrimination task \citep[e.g.,][]{Uka2003,Nienborg2009}, or area S1 in the context of a tactile discrimination task \citep[e.g.,][]{Hernandez2000}. We describe the activity of this neural population on every trial as a point process $\bs(t)=\{s_i(t)\}_{i=1\dots\Ntot}$, where each $s_i(t)$ is the spike train for neuron $i$, viewed as a series of Dirac pulses. As an important remark, $\Ntot$ denotes the full population size, a very large and unknown number. It is {\it not} the number of neurons actually recorded by the experimenter, which is generally much smaller.

For simplicity, we assume a fine discrimination task, where the different stimulus values $f$ presented to the animal display only moderate variations around a central value, say $f=0$. This substantially simplifies SNR computations, because the `signal' part of any response $r(f)$ is then summarized by its slope in $f=0$ : $\partial_f \pE(r|f)_{|f=0}$, where $\pE()$ denotes the average response over trials. We assume that this linearization with $f$ can be performed both for the psychometric report $\fstar$, and for individual neuron activities. This is mostly a convenience though, and the framework could be generalized to more complex dependencies on stimulus $f$.

From the raw data of $\fstar$ and $\bs(t)$ on each trial, a number of measures are routinely used to describe neural activity and animal behavior. First, the psychometric sensitivity $Z^\star$ describes the animal's accuracy in distinguishing nearby frequency values. It can be computed from the distribution of $(f,\fstar)$ across trials (\reffig{frame}b), according to the formula:
\begin{equation}
\label{eq:Zstar}
Z^\star=\frac{1}{\langle\Var(\fstar|f)\rangle_f},
\end{equation}
where notation $\langle.\rangle_f$ denotes an average across stimulus conditions. This is exactly the (squared) SNR for random variable $\fstar(f)$, assuming that the `signal' term $\partial_f \pE(\fstar|f)$ is equal to 1 because the animal's average judgement of $f$ is unbiased (the framework easily generalizes to a biased percept).

On the other hand, for each recorded neuron, it is common practice to compute its peri-stimulus time histogram (PSTH) in response to each different tested stimulus (\reffig{frame}d):
\begin{equation}
\label{eq:psth}
\lambda_{i}(t\;;f):=\pE(s_i(t)|f),
\end{equation}
where $\pE$ denotes averaging over trials. Since all stimuli $f$ are assumed to be close one from another, the dependency of $\lambda_{i}(t\;;f)$ on $f$ is essentially linear, and can be summarized by the (temporal) tuning curve for the neuron (\reffig{frame}e):
\begin{equation}
\label{eq:tuning}
\beta_{i}(t):=\partial_f \lambda_{i}(t\;;f).
\end{equation}

Furthermore, as recent techniques allow the simultaneous recording of many neurons, experimenters also have access to samples from the trial-to-trial covariance structure in the population (\reffig{frame}c). For every pair of neurons $(i,j)$ and instants in time $(t,s)$, this covariance structure is assessed through the neurons' joint peri-stimulus time histogram \citep[JPSTH,][]{Aertsen1989}:
\begin{equation}
\label{eq:jpsth}
\gamma_{ij}(t,s):=\langle\Cov(s_i(t),s_j(s)|f)\rangle_f.
\end{equation}
We only consider the average covariance structure, over different stimuli $f$. First, as above, nearby values of $f$ insure that the covariance structure will remain mostly unchanged. Second, trial-to-trial covariances correspond to second-order effects on neural activity, which require several trials to be reliably estimated---another reason to lump data from different stimuli $f$ into a single estimate.

Finally, we can measure a {\it choice signal} for each neuron, estimating the trial-to-trial covariance of neuron activity $s_i(t)$ with the animal's choice (\reffig{frame}f). Since in our framework the animal directly reports its percept $\fstar$, we readily describe the choice signal of each neuron by its {\it percept covariance} (PCV) curve:
\begin{equation}
\label{eq:pcov}
\pistar_{i}(t):=\langle\Cov(s_i(t),\fstar|f)\rangle_f.
\end{equation}
Again, this covariance information is lumped across the different (nearby) stimulus values $f$, in order to improve experimental measurement. The PCV curve captures the core intuition behind the more traditional measure of choice probability (CP), while retaining a linear form convenient for analytical treatment. Percept covariance curves are not directly measurable in classic experimental setups where the animal only reports a binary choice~;~however their analytical link to available measures such as CPs can be easily derived given simple models of the animal's decision policy (see \refsec{CP}).

Unlike many characterizations of neural activity that rely only on spike counts, our framework requires an explicit temporal description of neural activity through PSTHs (\refeq{psth}), JPSTHs (\refeq{jpsth}) and percept covariance curves  (\refeq{pcov}). Indeed, our method ultimately predict {\it when}, and {\it how long}, perceptual integration takes place in the organism. Readers may feel uncomfortable that the resulting definitions are directly expressed over trains of Dirac pulses. While these notations are fully justified in the framework of point processes \citep{Daley2007}, they describe idealized quantities that cannot be estimated from a finite number of trials, leading to jaggy estimates formed from the collection of Dirac peaks. So in practice, spike trains $s_i(t)$ are computed in temporal bins of finite precision.

\subsection{The readout model and its characteristic equations}
\label{sec:readout}

All experimental measures described above, taken together, provide a full characterization of the joint covariance structure of variables $(\bs(t),\fstar)$ across stimuli and trials (\reffig{model}c). The key argument to exploit these data, which is actually a reformulation of the `sensory noise' hypothesis, is that the animal's percept $\fstar$ is built on every trial from the activity of the sensory neurons, meaning that $\fstar=F^\star(\bs)$ for some unknown readout $F^\star$. As a result, each proposed readout $F$ directly yields an estimate for the joint covariance structure of $(\bs(t),\fstar)$---through a set of relationships which constitute the readout's {\it characteristic equations}. Conversely, since this joint covariance structure is experimentally measurable, it implicitly constrains the nature of the true readout $F^\star$ which was applied by the animal. In this section, we introduce a generic form of linear readout, stemming from the standard feedforward model of perceptual integration, and derive its characteristic equations. We show that in theory, these equations totally characterize the readout applied by the animal.

\subsubsection{Readout model}

\begin{figure}
\centerline{
         \includegraphics[width=0.9 \columnwidth]{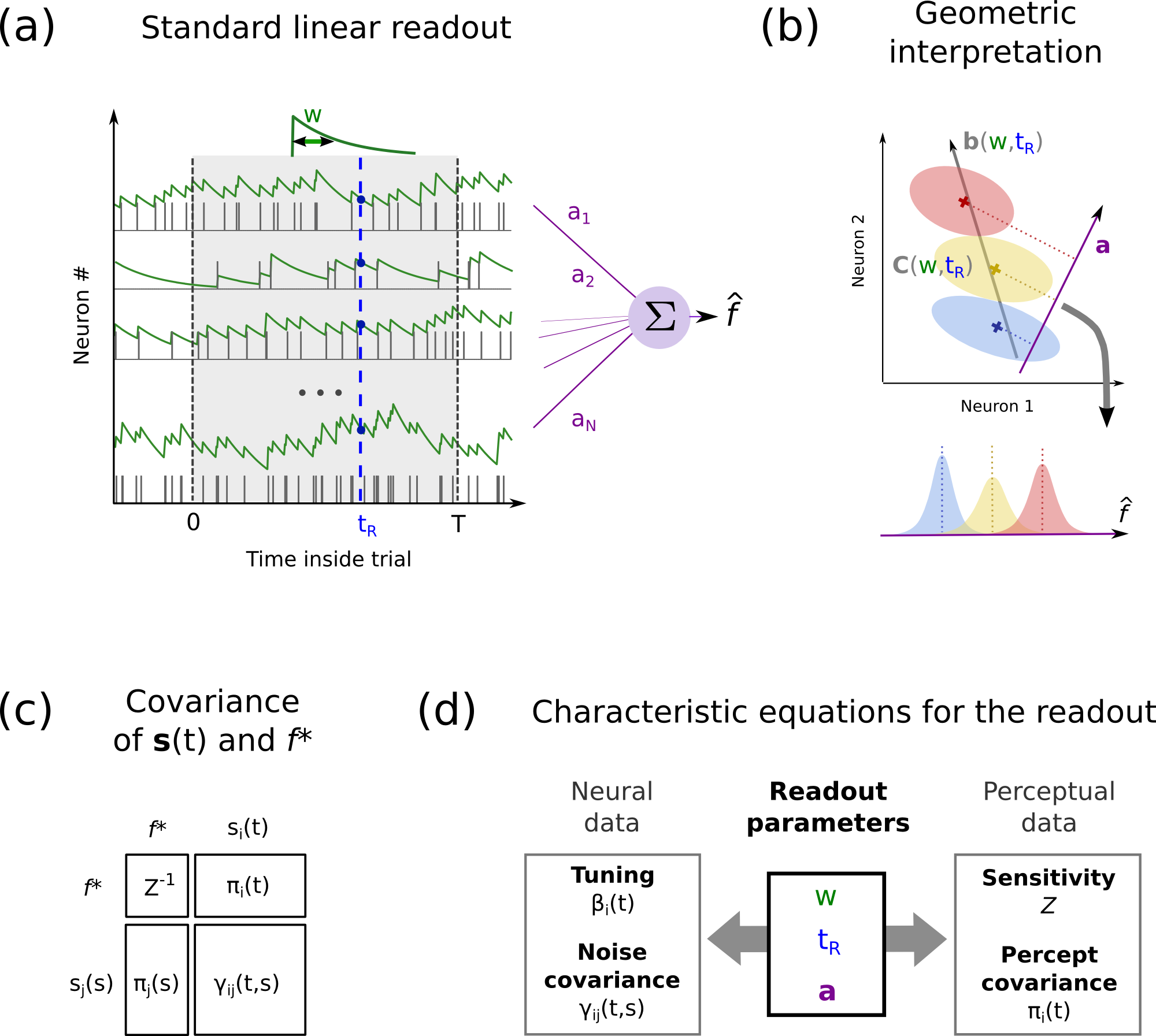}
	}
        \caption{\label{fig:model} Linear readout and its interpretation. (a) We study a ``standard'' model of perceptual readout, with two parameters $w$ and $\tR$ defining integration in time, and a readout vector $\ba$ defining integration across neurons. (b) Geometric interpretation of the model. The temporal parameters $w$ and $\tR$ define the tuning vector $\bb$ and noise covariance matrix $\bC$ in the population. Colored ellipses schematize the distribution of neural activities from trial to trial, for the three possible stimulus values. The readout $\fhat$ can be viewed as an orthogonal projection of neural activities in the direction given by $\ba$. (c) Sensitivity $Z$, PCV curves $\pi_i(t)$ and noise covariance JPSTHs $\gamma_{ij}(t,s)$ totally define the joint covariance structure between spike trains $\bs(t)$ and percept $\fhat$. (d) Any feedforward readout of neural activities can be viewed as a mapping $\fhat=F(\bs(t))$, so the true $F$ is implicitly constrained by the covariance data from panel c. In the case of the linear readout model, these constraints are summarized by three characateristic equations, which relate neural and perceptual data through the readout's parameters $w$, $\tR$ and $\ba$.}
\end{figure}

We define a generic linear readout from the activity of sensory neurons $\bs(t)$ (\reffig{model}a), based on a given readout vector: $\ba=\{a^i\}_{i=1\dots\Ntot}$, a given integration kernel with normalized shape $h$ and time constant $w$: $h_w(t):=w^{-1}h(\tau/w)$, and a given readout time $\tR$:
\begin{equation}
\label{eq:readout}
\fhat(\tR) := \sum_i \int_{u>0} a^i s_i(\tR-u) h_w(u) du.
\end{equation}
The readout is noted $\fhat$, as it must ultimately be an estimator of stimulus value $f$. We explicitely note the dependence on $\tR$ to emphasize that $\fhat(\tR)$ is built from a sliding temporal average of the spike trains~; so that each instant in time yields a potential readout.

This is a classical form of readout from a neural population, which has often been used previously and described as the `standard' model of perceptual integration \citep{Shadlen1996,Haefner2013}. The temporal parameters $h_w$ and $\tR$ describe how each neuron's temporal spike train $s_i(t)$ is integrated into a single number describing the neuron's activity over the trial: $\overline{s_i}=\int_{u>0} s_i(\tR-u)h_w(u)du$. In turn, the percept is built linearly from the population activity as $\fhat=\sum_i a^i\overline{s_i}$ through a specific readout vector, or `perceptual policy', $\ba$.

However, traditional studies generally make ad hoc choices for the various constituants of this readout. Most often, $\overline{s_i}$ simply describes the total spike count for neuron $i$, which in our model corresponds to choosing a square kernel $h$, and parameters $w=\tR=T$ describing an integration over the full period $T$ of sensory stimulation. As mentionned in the introduction, there is no reason that this should be a relevant description of sensory integration by the organism: the integration window $w$ has a direct influence on predicted SNRs for the readout, and experiments suggest that animals do not always use the full stimulation period to build their judgement \citep{Luna2005,Stanford2010}.

Instead, we make no assumption on the nature of $w$ and $\tR$, and view them as free parameters of the model. Then, the model parameters implicitly characterize the typical scales of perceptual integration by the animal. The number of significantly nonzero entries in $\ba$, say $\nK$, defines the number of neurons contributing to the percept. The readout window $w$ characterizes the behavioral scale of temporal integration from the sensory neurons, and time $\tR$ characterizes when during stimulation this integration takes place. The exact shape $h$ given to the integration kernel is of less importance~; for conceptual and implementational simplicity we assume it to be a square window. However, we note that (1) other shapes may have a higher biological relevance, such as the decreasing exponential mimicking synaptic integration by downstream neurons, and (2) nothing prevents our method from making $h$ itself a free parameter, provided the data contain enough power to estimate it. Finally, our model can also be extended to versions where extraction time $\tR$ is not fixed, but varies from trial to trial~; this issue is discussed in \refsec{withg}.

\subsubsection{Characteristic equations for the readout}
\label{sec:characteristic}


Thanks to its linear structure, the readout defined in \refeq{readout} allows for a simple characterization of the covariance structure that it induces between neural activity $\bs(t)$ and the resulting percept $\fhat$ (\reffig{model}b). We show in appendix \ref{sec:appChar} that this covariance structure can be summarized by three characteristic equations:
\begin{align}
1 &= \bb^\top\ba, \label{eq:char-tuning}\\
Z^{-1} &= \ba^\top\bC\ba, \label{eq:char-Z}\\
\bpi(t)&= \bGam(t)\ba, \label{eq:char-pcov}
\end{align}
where vector $\bb$ and matrices $\bGam(t)$ and $\bC$ respectively describe the population's tuning and noise covariance structures, derived from the underlying neural statistics $\bbeta(t)$ and $\bgam(t)$ introduced in \refeq{tuning}-\ref{eq:jpsth}:
\begin{align}
b_i(w,\tR) &:= \int_{u>0} \beta_i(\tR-u)h_w(u)du,\label{eq:b}\\
\Gamma_{ij}(t\,|w,\tR) &:= \int_{u>0} \gamma_{ij}(t,\tR-u)h_w(u)du,\label{eq:Gamma}\\
C_{ij}(w,\tR) &:= \int_{u>0} \Gamma_{ij}(\tR-u)h_w(u)du.\label{eq:C}
\end{align}
We here note the explicit dependency of $\bb$, $\bGam(t)$ and $\bC$ on the temporal parameters of the readout $w$ and $\tR$. We will generally omit it in the sequel. Thus, the right-hand sides of \refeq{char-tuning}-\ref{eq:char-pcov} depend only on readout parameters $w$, $\tR$, $\ba$ and on the statistics of neural activity, independently of the animal's percept.

On the other hand, the left-hand sides of \refeq{char-tuning}-\ref{eq:char-pcov} describe experimental quantities related to the readout's resulting percept $\fhat$. The first line describes the average tuning of $\fhat$ to stimulus $f$, that is $\partial_f \pE(\fhat|f)$, which is equal to 1 because we assume that $\fhat$ is unbiased. The second line expresses the resulting sensitivity $Z$ for the readout, defined as in \refeq{Zstar}. It reveals the dual influence of the number of neurons (through $\ba$) and integration window $w$ on the readout's overall sensitivity: indeed, under mild assumptions, the covariance matrix $\bC$ scales with $w^{-1}$ (see appendix \ref{sec:appw}). 
Finally, the third line expresses the resulting covariance between $\fhat$ and the activity of each neuron $s_i(t)$, defined as in \refeq{pcov}. This is essentially the relationship already revealed by \citet{Haefner2013}, that choice probabilities are related to readout weights through the noise covariance matrix~; however, our formalism focuses on the simpler linear measure of PCV curves, and explicitly takes time into account.

Both the neural measures $\bbeta(t)$ and $\bgam(t,s)$ on the right-hand side, and the percept-related measures $Z$ and $\bpi(t)$ on the left-hand side, can be estimated from data. As a result, the characteristic equations define an implicit constraint on the readout parameters $w$, $\tR$ and $\ba$ (\reffig{model}d). Actually, if the readout model in \refeq{readout} is true, and precise measures are available for all neurons in the population, one sees easily that these constraints would uniquely determine the readout parameters. Indeed, for fixed parameters $w$ and $\tR$, \refeq{char-tuning} and \ref{eq:char-pcov} impose linear constraints on vector $\ba$. These constraints are generally overcomplete, since $\ba$ is $\Ntot$-dimensional, while each time $t$ in \refeq{char-pcov} provides $\Ntot$ additional linear constraints. Thus, in general, a solution $\ba$ will only exist if one has targeted the true parameters $w$ and $\tR$, and it will then be unique.

\subsection{Estimating the scales of sensory integration}
\label{sec:statistical}

In the previous section we have shown that, in the standard linear model of percept formation, the trial-to-trial covariance structure between spike trains $\bs(t)$ and the resulting percept $\fhat$ leads to a set of characteristic equations which implicitly define the parameters of the perceptual readout, provided the covariance structure has been fully estimated.

Unfortunately, this direct approach makes a fundamental assumption which cannot be reconciled with real, experimental recordings: it assumes we have recorded all neurons from the population under survey, whereas real recordings only ever record from a small subset of that population. Thus we cannot hope to reconstruct the real vector $\ba$, simply because some---probably most---of the neurons contributing to $\ba$ were not recorded. Moreover, even across those neurons which were recorded through a series of sessions in a given area, the noise covariance structure can never be fully assessed~; it remains elusive between neurons which were not recorded simultaneously.

For this fundamental reason, the characteristic equations \ref{eq:char-tuning}-\ref{eq:char-pcov} should be used with a different perspective than the full recovery of readout parameters. Instead, we propose to exploit the structure of the equations in a statistical approach, with the restricted goal of estimating the typical scales of readout most compatible with recorded data.

\subsubsection{Reformulation in terms of neural subensembles}

A first necessary step in our approach is to statistically describe the nature of readout vector $\ba$. We are mostly interested in the support of $\ba$, meaning, the number and nature of neurons contributing to percept formation. Thus, we assume that the percept is built only from the activities of an unknown ensemble $\Kap$ of neurons in the population and that, for given $\Kap$ and temporal parameters $(w,\tR)$, the readout vector $\ba$ is chosen optimally to maximize the SNR of the resulting percept. Indeed, through this hypothesis, we totally reformulate the problem of characterizing $\ba$ in that of characterizing $\Kap$~; which allows for much simpler statistical descriptions.

The readout vector $\ba$ achieving the maximum sensitivity $Z$ in \refeq{char-Z}, under the constraints of \refeq{char-tuning} and having support on $\Kap$, is well known from the statistical literature. It is uniquely given by Fisher's linear discriminant formula \citep{Hastie2009}:
\begin{align}
\ba_\Kap&= \frac{1}{\bb_\Kap^\top \bC_\Kap^{-1} \bb_\Kap^{~}}\;\bC_\Kap^{-1} \bb_\Kap^{~}, \label{eq:a-opt}
\end{align}
where $\ba_\Kap$, $\bb_\Kap$ and $\bC_\Kap$ are the versions of vectors $\ba$, $\bb$ (\refeq{b}) and matrix $\bC$ (\refeq{C}) restricted to neuron ensemble $\Kap$. By injecting the form (\refeq{a-opt}) into \refeq{char-Z}-\ref{eq:char-pcov} we obtain a new version of the characteristic equations, under the assumption that percept is built optimally from some given ensemble $\Kap$, and temporal parameters $(w,\tR)$:
\begin{align}
Z(\Kap\;|w,\tR) &= \bb_\Kap^\top \bC_\Kap^{-1} \bb_\Kap^{~}, \label{eq:new-Z} \\
\pi_i(t\;|\Kap,w,\tR) &= \frac{1}{Z(\Kap)}\;\bGam_{i \Kap}^{~}(t) \bC_\Kap^{-1} \bb_\Kap^{~}.\label{eq:new-pcov}
\end{align}
$Z$ in \refeq{new-Z} is the (optimal) sensitivity associated to this particular choice of $\Kap$, $w$ and $\tR$. In \refeq{new-pcov}, $\pi_i(t)$ is the resulting, predicted PCV curve for every neuron $i$ in the population (not only in ensemble $\Kap$). $\bGam_{i \Kap}(t)$ is a row vector whose entries are equal to $\Gamma_{ij}(t)$ (\refeq{Gamma}) for neurons $j \in \Kap$.\\

These equations open the door to a statistical description of percept formation in the neural population: we can now parse through a large set of candidate ensembles $\Kap$ and temporal parameters $(w,\tR)$, and ask when the predictions for sensitivity (\refeq{new-Z}) and PCV curves (\refeq{new-pcov}) match their true psychophysical counterparts $\Zstar$ (\refeq{Zstar}) and $\pistar_i(t)$ (\refeq{pcov}). For sensitivity, the straightforward comparison is to require that $Z(\Kap\;|w,\tR)\approx\Zstar$.

On the other hand, for the PCV equation (\refeq{new-pcov}), it is pointless to search an elementwise match for every neuron $i$, between the predicted curve $\pi_i(t)$ and its true measure $\pistar_i(t)$. Indeed, since only a small subset of the neurons have been recorded, no candidate readout ensemble $\Kap$ will be equal to the true ensemble (say $\Kapstar$) that was used by the animal~; and there is no guarantee that the covariance structure between $i$ and $\Kap$, which gives rise to prediction (\refeq{new-pcov}), should be similar to that between $i$ and $\Kapstar$. Instead, a given set of readout parameters $(\Kap,w,\tR)$ should be deemed plausible if they predict the correct {\it distribution} of PCV signals across the population, irrespective of exact neuron identities $i$. Full distributions are difficult to estimate from finite amounts of data, and we will find the following population {\it averages} to convey sufficient information:
\begin{align}
W(t\;|\Kap,w,\tR) &:= \pE_i \Big(b_i(w,\tR) \pi_i(t\;|\Kap,w,\tR)\Big),\label{eq:W-kap}\\
\Wstar(t\;|w,\tR) &:= \pE_i \Big(b_i(w,\tR) \pistar_i(t)\Big),\label{eq:Wstar}
\end{align}
where $\pE_i$ denotes averaging over the full population of neurons $i=1\dots \Ntot$. We will deem a set of readout parameters plausible if they yield $W(\Kap\;|w,\tR)\approx\Wstar(w,\tR)$\footnote{Note that $\Wstar(t\;|w,\tR)$ depends on parameters $(w,\tR)$ only through the neurons' tunings $b_i(w,\tR)$. In practice, as neural activities are rather stationary in time, $\Wstar(t)$ changes very little for different values of parameters $(w,\tR)$.}. Multiplying each PCV curve by the neuron's tuning $b_i$ (\refeq{b}) yields more stable estimates for $W$, as discussed in \refsec{full-valid} and appendix \ref{sec:appSVD}.

\subsubsection{Statistical constraints on readout scales}

There are many ways to compare the real values of sensitivity and PCV signals, to their predictions given by \refeq{new-Z}-\ref{eq:new-pcov}. We propose here an ad-hoc method, whose main characteristics are the following: (1) focus mostly on first-order statistics (i.e., means) across the neural population, (2) use arbitrary tolerance values to compare real and predicted data, (3) fit the two indicators sequentially: first SNR, then percept covariance. Due to its simplicity, this method will prove robust to measurements errors arising from finite amounts of data (\refsec{finite}).

Our method is also designed to cope with a fundamental limitation of real recordings: all neurons (ensemble $\Kap$, neurons $i$) contributing to predictions \refeq{new-Z}-\ref{eq:new-pcov} must have been recorded simultaneously, to assess their noise covariance structure. This constraint sets a limit on ensemble sizes $K$ which can be easily investigated (but see \refsec{extrapolation}). Moreover, it prevents from estimating the full average of choice signals (\refeq{W-kap}) predicted by a given ensemble $\Kap$---it is only available for simultaneously recorded neurons $i$. As a result, predictions (\refeq{new-pcov}) from different tested ensembles $\Kap$ must somehow be aggregated to produce a reliable prediction of choice signals.

We propose that each tested ensemble $\Kap$ contribute to our estimates in proportion to its ability to account for the animal's sensitivity:
\begin{equation}
\PSNR(\Kap\;|w,\tR)\sim \mathrm{exp}\left(-\frac{(Z(\Kap\;|w,\tR)-\Zstar)^2}{2\aSNR^2}\right),\label{eq:Psnr}
\end{equation}
normalized to insure $\sum_\Kap \PSNR(\Kap)=1$ across all tested ensembles ($w$ and $\tR$ being fixed). Parameter $\aSNR$ is the required tolerance for the fit, set by the experimenter. It is a regularization parameter creating a tradeoff between precision of fit (small $\aSNR$) and reliability of measurements, since a larger $\aSNR$ leads to more samples $\Kap$ with a substantial contribution $\PSNR(\Kap)$. When testing our method (\refsec{results}) we choose $\aSNR$ as 5\% of $\Zstar$.

For each tested couple $(w,\tR)$, we then use $\PSNR(\Kap\;|w,\tR)$ as a weighting factor over all tested ensembles $\Kap$, which yields two quantities:
\begin{align}
\Kest(w,\tR) &:= \sum_\Kap \PSNR(\Kap\;|w,\tR)\mathrm{Card}(\Kap),\label{eq:Kest}\\
\West(t\;|w,\tR) &:= \sum_\Kap \PSNR(\Kap\;|w,\tR) \pE_{i(\Kap)} \Big(b_i \pi_i(t\;|\Kap,w,\tR)\Big),\label{eq:West}
\end{align}
where $\pE_{i(\Kap)}$ denotes an average across all neurons $i$ available to compute a prediction with \refeq{new-pcov}. These neurons must have been recorded simultaneously to ensemble $\Kap$ and, in order to produce an unbiased estimate of choice signals in the full population, they should not belong to $\Kap$ itself.

In \refeq{Kest}, $\Kest(w,\tR)$ is the ensemble size $K$ which most likely explains the animal's sensitivity, given readout parameters $(w,\tR)$. In \refeq{West}, $\West(t\;|w,\tR)$ is the mean prediction for PCV signals $b_i \pi_i(t)$ across neurons $i$ in the population, but stemming only from ensembles $\Kap$ which are compatible with the animal's sensitivity. Considering quantity $W(t)$ introduced in \refeq{W-kap}, we see that
\begin{equation}
\label{eq:Wapp}
\West(t\;|w,\tR) \simeq \sum_\Kap \PSNR(\Kap\;|w,\tR) W(t\;|\Kap,w,\tR).
\end{equation}
The equality is only approximate, because only neurons $i$ recorded simultaneously to $\Kap$ are available to estimate $\West(t)$. However, as neurons $i$ are random and we average over many ensembles $\Kap$, $\West(t)$ rapidly converges to the quantity described in \refeq{Wapp}. 

Both $\West(t\;|w,\tR)$ and $\Wstar(t\;|w,\tR)$ are temporal signals defined over some interval $[T_\mathrm{min},T_\mathrm{max}]$ corresponding to one trial repetition. Defining the L2 norm for such temporal signals as 
\begin{equation}
\label{eq:L2}
\|x\|^2=(T_\mathrm{max}-T_\mathrm{min})^{-1}\int_{t=T_\mathrm{min}}^{T_\mathrm{max}}x^2(t)dt,
\end{equation}
we will deem parameters $(w,\tR)$ plausible if they lead to a small value of $\|\West(w,\tR)-\Wstar(w,\tR)\|^2$. To yield a quantitative estimate of fit, we introduce a tolerance $\aPCV$ and define the following weighting function:
\begin{equation}
\PPCV(w,\tR)\sim \mathrm{exp}\left(-\frac{\|\West(w,\tR)-\Wstar(w,\tR)\|^2}{2\aPCV^2}\right),\label{eq:Ppcv}
\end{equation}
normalized to insure $\sum_{w,\tR} \PPCV(w,\tR)=1$ across all tested temporal parameters $(w,\tR)$. Again, tolerance $\aPCV$ is set arbitrarily by the experimenter. When testing our method (\refsec{results}) we choose $\aPCV$ as 5\% of $\|\Wstar(w,\tR)\|$.\\

Overall, the statistical method introduced above reduces readout parameters to three numbers: the temporal extraction parameters $w$ and $\tR$, and the typical number of neurons $K$ used by the readout. Thus, we can now apply a `brute-force' approach: test all possible combinations $(K,w,\tR)$, compute the population statistics from \refeq{Psnr}-\ref{eq:Ppcv}, and target the parameters that provide the best fit. In the next section, we show the validity of this statistical approach, which allows us to recover the typical scales $(K,w,\tR)$ of perceptual integration in an artificial network simulation. We further detail how this statistical approach can be adapted to counteract measurement errors which typically arise in real experiments from the finite number of available trials.

%
%

\section{Results}
\label{sec:results}

\subsection{Artificial neural network}
\label{sec:network}

In this section, we show how the statistical analysis of sensitivity and choice signals described above allows to recover the scales of integration of the neural readout. Naturally, to assess the validity of our method, it is necessary to know the true nature of this readout. This can only be achieved through an artificial simulation of sensory integration, where we have full control on neural activities and readout procedure.

\begin{figure}
\centerline{
         \includegraphics[width=.9 \columnwidth]{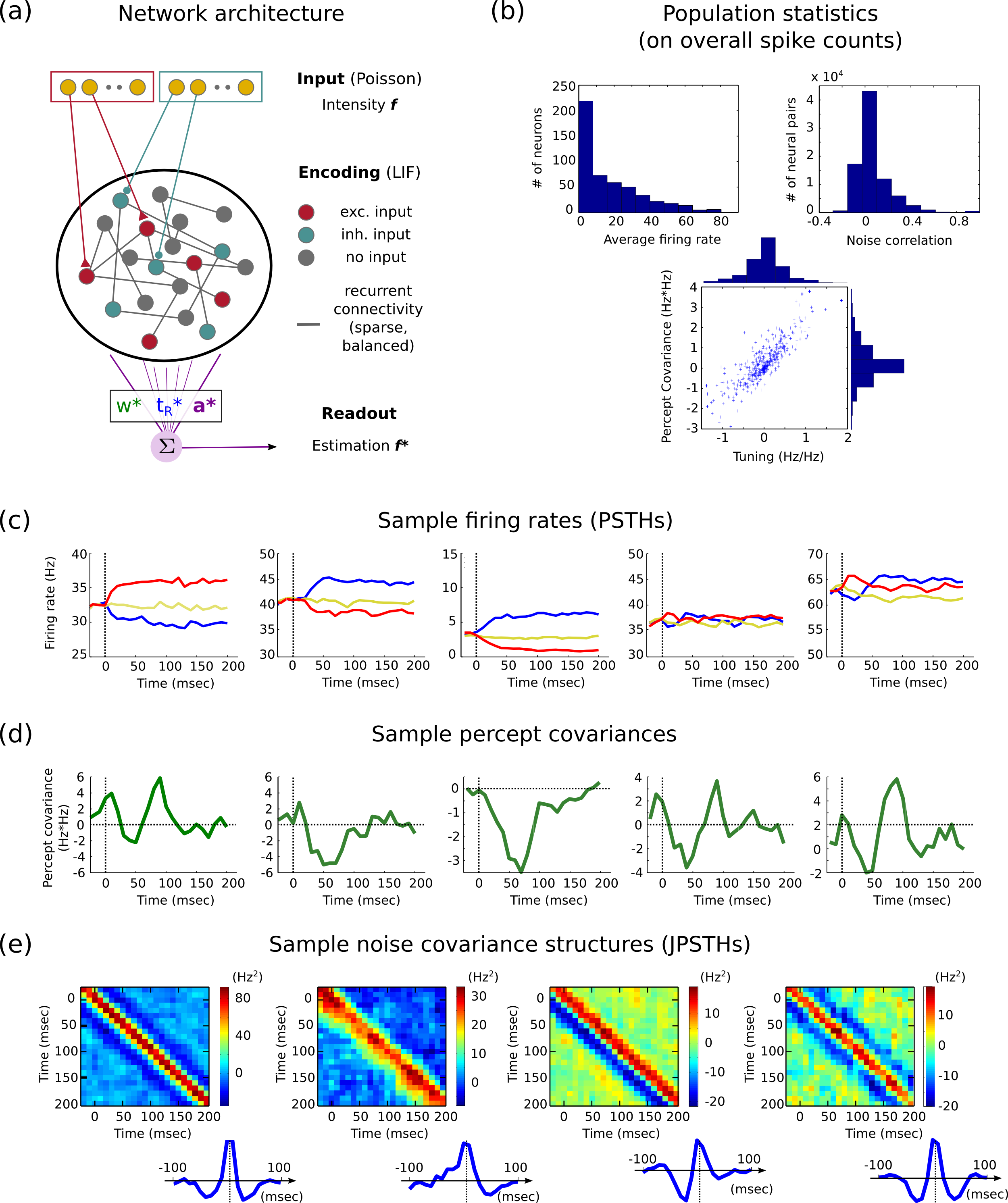}
	}
        \caption{\label{fig:network} Artificial neural network used for testing the method. (a) Network architecture. The encoding layer consists of LIF neurons coupled through a sparse, balanced recurrent connectivity with random delays. Stimulus $f$ is the firing intensity of a group of input Poisson neurons, which project sparsely into two subpopulations of the encoding layer (neurons excited by the input vs. neurons inhibited by the input). The majority of neurons in the encoding layer receive no direct projection, but can still acquire stimulus tuning through the recurrent connections. A ``true'' readout $\fstar$ is produced on every trial on the basis of true parameters $\wstar$, $\tRstar$ and $K^\star$---which should be retrieved by our method. (b) Classic population statistics in the encoding layer, for neural spike counts over a trial. (c) Sample PSTHs from the encoding layer. Model neurons display varied firing rates, and tunings of different polarities. (d) Sample PCV curves for the same neurons as panel c, computed by correlating each neuron's spikes with the true readout $\fstar$. (e) Sample JPSTHs (noise correlations) for pairs of neurons in the encoding layer. Inset: corresponding cross-correlograms, obtained by projection along the diagonal.}
\end{figure}

We thus implemented an artificial neural network, that encodes some input stimulus $f$ in the spiking activity of its neurons (\reffig{network}a). Precise parameters of this network are provided as Supplementary Material (section S1). 
Briefly, on each trial, 100 input Poisson neurons fire with rate $f$, taking one of three possible values 25, 30 and 35 Hz. The encoding population {\it per se} consists of 500 leaky integrate-and-fire (LIF) neurons. 100 of these neurons receive sparse excitatory projections from the input Poisson neurons, which naturally endows them with a positive tuning to stimulus $f$. 100 other neurons receive sparse inhibitory projections from the Poisson neurons, which naturally endows them with negative tuning. The remaining 300 neurons receive no direct projections from the input. Instead, all neurons in the encoding population are coupled through a sparse connectivity with random delays up to 5 ms. Synaptic weights are random and balanced, tuned to ensure overall firing rates around $30$ Hz. We implemented and simulated the network using Brian, a spiking neural network simulator in Python \citep{Goodman2008}. The statistics of activity for the resulting population are depicted in \reffig{network}b,c,e.

We then define the true perceptual readout from this network. We pick a random set of $K^\star=40$ neurons in the population, whose activity is integrated over $\wstar = 50$ ms and read out at time $\tRstar=80$ ms, on each stimulus presentation (each presentation lasting 500 ms). The resulting estimator $\fstar$ of stimulus value is built optimally given these constraints, through Fisher linear discriminant analysis (\refeq{a-opt})\footnote{To avoid overfitting issues, the trials used to learn the optimal $\fstar$ are not used in the subsequent analysis.}. This leads to a `psychometric' sensitivity $\Zstar \approx$ 0.06 Hz$^{-2}$, meaning that the network can typically discriminate variations of $(\Zstar)^{-1/2}\approx$ 4.2 Hz in the input $f$. (For comparison, over the same integration period $\wstar$, the 100 input Poisson neurons can discriminate variations around 2.5 Hz.)
We then compute a PCV for every recorded neuron, measuring its trial-to-trial covariance with estimator $\fstar$ (\reffig{network}d).  Applying the statistical method described above, our goal is now to recover the scales $(K^\star,\wstar,\tRstar)$ of perceptual integration, on the basis of the experimental measures depicted in (\reffig{network}c-e).

\subsection{Validation on full population data}
\label{sec:full-valid}

To show the theoretical validity of our analysis, we first apply it to a situation where the trial-to-trial covariance of neurons and percept $\fstar$ (\refeq{Zstar}-\ref{eq:pcov}) has been fully measured, with high precision\footnote{To compute these estimates, all 500 neurons were simultaneously monitored over 16,500 repetitions of each stimulus condition (not counting the trials used to train estimator $\fstar$). Using bootstrap resamplings over stimulus repetitions, we checked that the resulting measures were virtually error-free.}. In particular, we assume full knowledge of the noise covariance structure $\bgam(t,s)$ in our population (\refeq{jpsth}). Actually, in these irrealistic conditions, the statistical analysis described above is not useful: instead, one could directly solve the characteristic equations (\refeq{char-tuning}-\ref{eq:char-pcov}) to recover the readout parameters $\ba$, $w$ and $\tR$. However, this is a necessary first step to verify that our method is not flawed theoretically. Do the statistical quantities introduced in \refeq{Psnr}-\ref{eq:Ppcv} allow to recover the true scales of integration $(K^\star,\wstar,\tRstar)$?

\begin{figure}
\centerline{
         \includegraphics[width=\columnwidth]{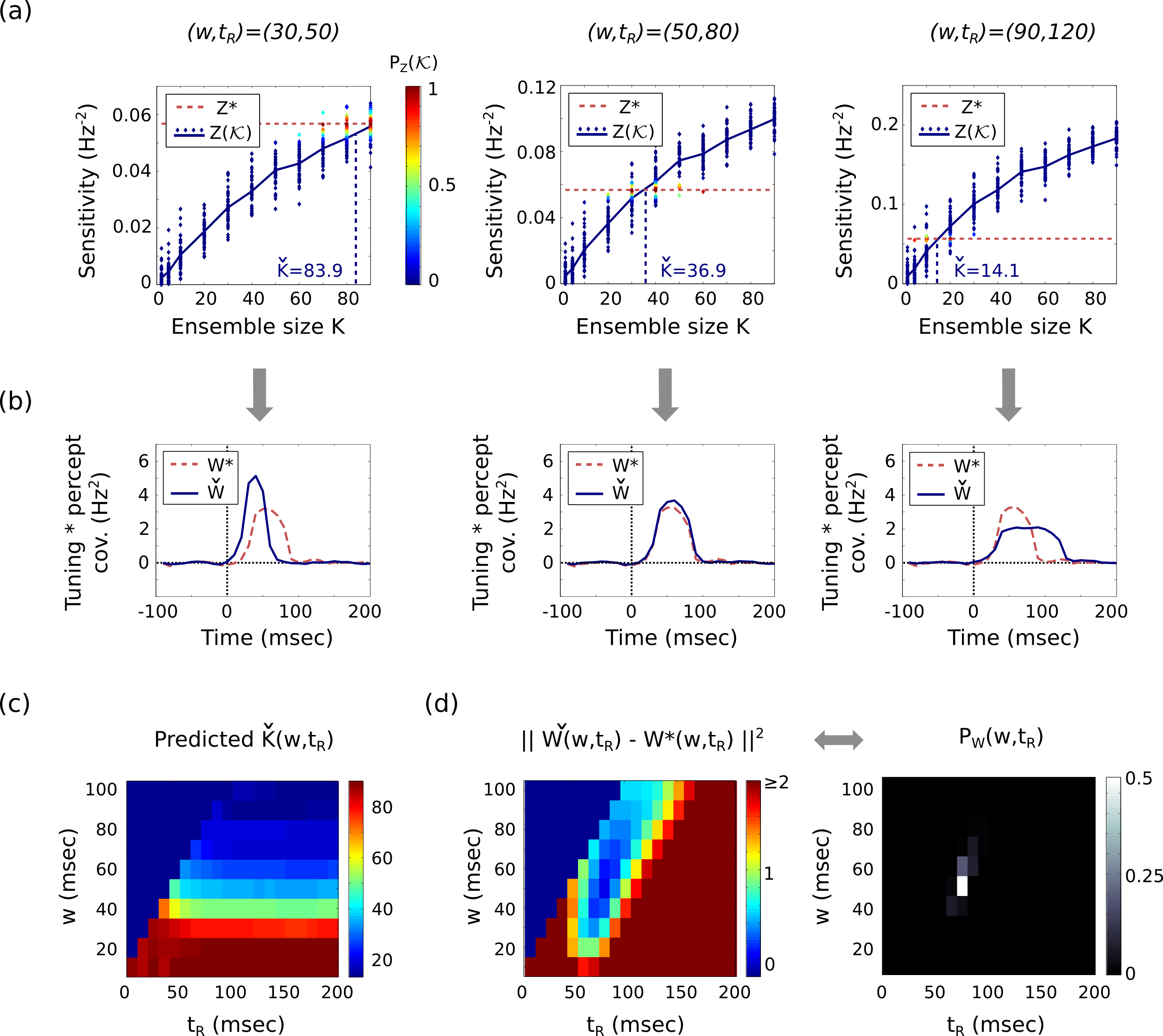}
	}
        \caption{\label{fig:noiseless} Statistical recovery of readout parameters: noiseless measures. (a) For each tested temporal parameters $(w,\tR)$, predicted sensitivities $Z(\Kap)$ are computed for several candidate readout ensembles $\Kap$ of varying sizes. The goodness of fit to true sensitivity $\Zstar$ defines a weighting function $\PSNR(\Kap)$ across ensembles. (b) The weighting function is used to compute a compound prediction $\West(t)$ for the average PCV signal in the population, which is compared to the true average $\Wstar(t)$. The three columns in panels a-b correspond to different candidates $(w,\tR)$ for temporal integration. (c) Best-fitting ensemble size $\Kest$ depending on candidate parameters $(w,\tR)$. The $K-w$ tradeoff on sensitivity is clearly visible. (d) Goodness of fit of PCV signals depending on candidate parameters $(w,\tR)$ shows a clear optimum around the true parameters of the readout. (e) Same as panel d, but transformed into a weighting function $\PPCV(w,\tR)$ over candidate temporal parameters.}
\end{figure}

Assuming a square integration kernel $h$, we test a set of candidate temporal integration windows $w$ from 10 to 100 msec, and a set of candidate readout times $\tR$ from 10 to 200 msec, all in steps of 10 msec. We then pick randomly candidate neural ensembles $\Kap$, of sizes ranging from 2 to 90 neurons, with 50 different random ensembles for each tested size $K$. For each tested parameters $(w,\tR)$, we compute the distribution of predicted SNRs $Z(\Kap)$ given by \refeq{new-Z}, across all candidate neural ensembles (\reffig{noiseless}a). Each neural sample $\Kap$ is then associated to a weight $\PSNR(\Kap)$ describing the goodness of fit to the true SNR $\Zstar$ (\refeq{Psnr}). Following \refeq{Kest}-\ref{eq:West}, this yields an estimate for the best-fitting population size $\Kest(w,\tR)$ (\reffig{noiseless}a, dashed vertical line) and mean PCV curve $\West(t\;|w,\tR)$ (\reffig{noiseless}b). Since we assume full knowledge of experimental data, all 500 neurons $i$ are involved in estimating $\West(t)$, independently of ensemble $\Kap$.

In \reffig{noiseless}c, we show the estimated population size $\Kest(w,\tR)$ as a function of $w$ and $\tR$. It shows the mark of the $K$--$w$ tradeoff on sensitivity, mentioned in the introduction: smaller integration windows $w$ require larger ensemble sizes $K$ to account for the animal's sensitivity. In \reffig{noiseless}d, we show two measures of the resulting fit between $\West(t\;|w,\tR)$ and its true value $\Wstar(t\;|w,\tR)$. In the first panel, we plot the plain L2 norm between the two temporal signals (using $T_\mathrm{min}=-100$ msec and $T_\mathrm{max}=200$ msec as integration bounds). In the second panel, we reexpress this L2 norm as a weighting $\PPCV(w,\tR)$ over the set of tested temporal parameters (\refeq{Ppcv}). Applying this final weighting over candidate values $w$, $\tR$, and $\Kest(w,\tR)$ yields numerical estimates for the scales of the readout:
\begin{align*}
\widehat{w}&=54 \pm 9 \;\;\textrm{msec}\\
\widehat{t_R}&=81 \pm 6 \;\;\textrm{msec}\\
\widehat{K}&=34.7 \pm 8.7
\end{align*}
These estimates are very close to the true values $\wstar$, $\tRstar$ and $K^\star$, showing the theoretical validity of this approach. The estimated $\widehat{K}$ is somewhat smaller than its true value $K^\star=40$, however this is no bias in our method: it simply means that the 40 neurons chosen randomly as the source of percept were slightly less sensitive than the `average' 40 neurons in the population.

We also remind that these estimates depend on the tolerance levels fixed by the experimenter to compute $\PSNR(\Kap)$ (\refeq{Psnr}) and $\PPCV(w,\tR)$ (\refeq{Ppcv}). Numerically, we find the resulting mean estimates to be rather stable across a range of sensible tolerances. On the other hand, the resulting error bars---which are obtained as second-order moment of the quantities weighted by $\PPCV(w,\tR)$---only describe the typical variations of the parameters that lead to estimates within the fixed tolerances. In particular, driving the tolerances to zero always drives the error bars to zero, even though the predicted averages may become false as too little data enter their computation.\\

Why does the method work? Essentially, it proceeds in two successive steps. First, $(w,\tR)$ being held fixed, it uses SNR information to target plausible neural ensembles $\Kap$ (\reffig{noiseless}a,c). Since the readout is assumed to be optimal, the mean SNR can only increase with the size $K$ of the ensembles considered (\reffig{noiseless}a, plain blue curve). Plausible ensembles $\Kap$ are those lying near the crossing of this curve with the true `psychometric' SNR (\reffig{noiseless}a, dashed red curve). For a straightforward application of our method, this crossing should occur within the typical ensemble sizes $K$ tested---which are, in practice, limited by the number of simultaneously recorded neurons. In \refsec{extrapolation}, we discuss possible extensions of the method to the case where the crossing does not occur.

\begin{figure}
\begin{center}
         \includegraphics[width=\columnwidth]{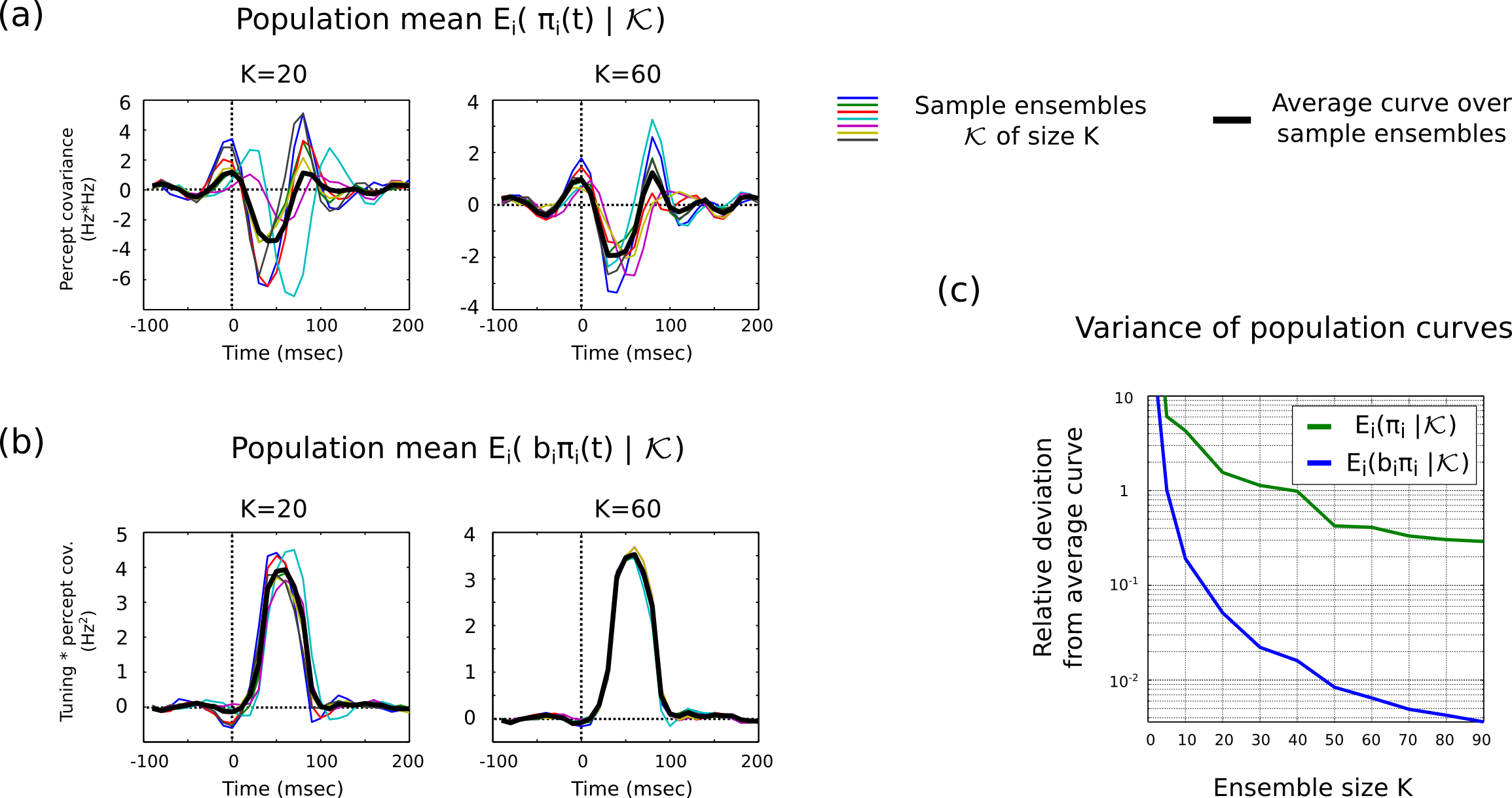}
\end{center}
        \caption{\label{fig:prod-pcov} Mean percept covariance curves depend on readout ensemble $\Kap$. (a) The mean value of PCV curves $\pi_i(t)$ across neurons $i$ in the population depends strongly on the readout ensemble $\Kap$ giving rise to the percept. (b) The mean value of tuning-multiplied PCV curves $b_i \pi_i(t)$ depends much less on the exact ensemble $\Kap$, only on its size. This justifies our definition for the mean PCV curve $W(t|\Kap)$ (\refeq{W-kap}). (c) Relative variance of mean PCV curves, across readout ensembles $\Kap$ of the same size. It is defined as the average of $\|W(\Kap_1)-W(\Kap_2)\|^2$ across all ensembles ($\Kap_1$,$\Kap_2$) of similar size, divided by $\|\pE_K W(\Kap)\|^2$, power of the average curve across ensembles of size $K$. For the tuning-multiplied version of $W(t|\Kap)$ (blue), this ratio quicky drops to zero. This is not the case for the plain mean $\pE_i(\pi_i(t)|\Kap)$ (green).}
\end{figure}

Second, $(w,\tR)$ being still fixed, an average PCV prediction $\West(t\;|w,\tR)$ is built, using the neural ensembles $\Kap$ targeted above, and compared to the true mean PCV curve $\Wstar(t\;|w,\tR)$. It is not trivial that this comparison should work. To simplify the argumentation, let us assume that parameters $w$ and $\tR$ are fixed at their true values $\wstar$ and $\tRstar$. On the one hand, since the true percept is built from some (unknown) neural ensemble $\Kapstar$, we have $\Wstar(t) = W(t\;|\Kapstar)$, using the notations of \refeq{W-kap}-\ref{eq:Wstar}. On the other hand, the prediction $\West(t)$ is built as a compound mean of $W(t\;|\Kap)$ over several candidate ensembles $\Kap$ (see \refeq{Wapp}). As our method requires a match between $\Wstar(t)$ and $\West(t)$, it implicitly supposes that all ensembles $\Kap$ contributing to $\West(t)$ lead to very similar population averages $W(t\;|\Kap)$.

Predicted curves $W(t\;|\Kap)$ are generally not available experimentally. However, we can compute them in our full-data simulation (\reffig{prod-pcov}). We find that, amongst ensembles $\Kap$ of similar size $K$, the $i$-population means of $b_i \pi_i(t\;|\Kap)$ rapidly converge to a single curve, independently of ensemble $\Kap$ (\reffig{prod-pcov}b). Furthermore, this result is not trivial: when the same analysis is performed on the plain PCV curves, not multiplied by tuning $b_i$, the convergence does not occur anymore (\reffig{prod-pcov}a), or at least not as fast. In \reffig{prod-pcov}c, we plot the ratio between the variance of curves accross different ensembles $\Kap$, and the power of the mean curve, across all ensembles $\Kap$ of same size $K$. This ratio quickly drops to zero for the tuning-multiplied PCV curves (blue), but not for the plain PCV curves (green).

To summarize, it is crucial for our method that each PCV curve $\pi_i(t)$ be multiplied by the neuron's tuning $b_i$ before computing population averages. Aside from the experimental observations of \reffig{prod-pcov}, several arguments justify this operation.
First, it is well-known experimentally that choice signals and tuning for individual neurons are often positively correlated at the population level \citep{Britten1996,Uka2003}. Intuitively, this is because positively-tuned neurons contribute positively to stimulus estimation, and conversely for negatively-tuned neurons.
The strong population-wide correlation is indeed present in our simulated network (\reffig{network}b). As a result, the population average for $b_i \pi_i(t)$ is expected to be mostly positive (\reffig{prod-pcov}b), which diminishes possible variations from one ensemble $\Kap$ to the other. Second, theoretical arguments (appendix \ref{sec:appSVD} and Supplementary Material S2) 
show that $b_i \pi_i(t)$ is a form better suited to compute an $i$-population average. It can be shown to be positive under mild assumptions, and its laws of convergence can be related to the overall spectrum of covariance in the population.

\subsection{Validation on finite data}
\label{sec:finite}

\begin{figure}
\centerline{
         \includegraphics[width=\columnwidth]{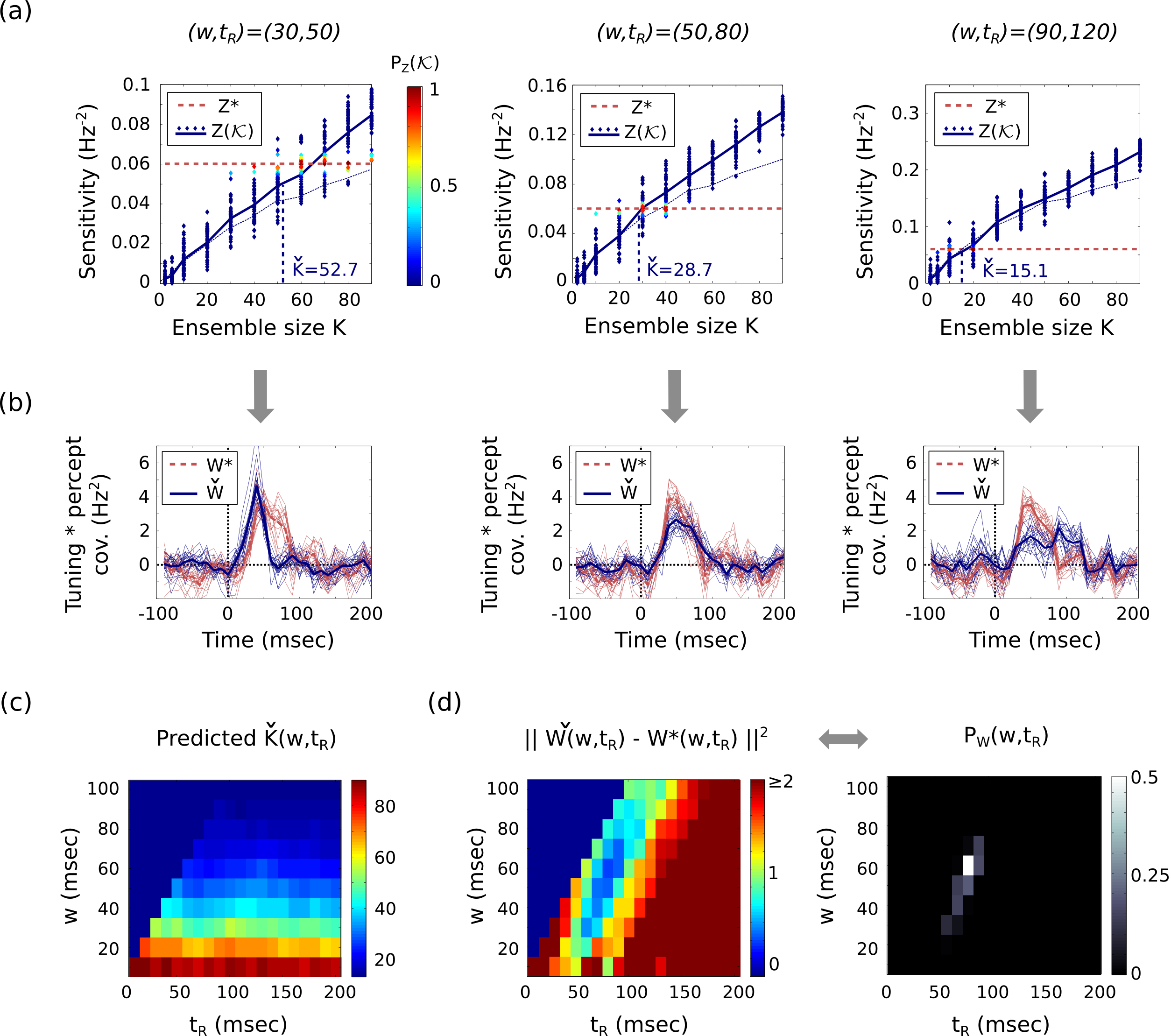}
	}
        \caption{\label{fig:noisy} Statistical recovery of readout parameters: noisy measures. Same legends as \reffig{noiseless}, but with modifications specific to small sample data. In panel b, the thin curves are different versions obtained through bootstrap resampling over trials, and the thick curve is the average across bootstrap samples. In panel d, the L2 norm is corrected for measurement errors, using the bootstrap samples and \refeq{Wboot}. With this modification, our method can recover the true readout parameters on the basis of finite amounts of data.}
\end{figure}

Having shown the theoretical efficiency of the statistical quantities introduced above in retrieving the correct scales of perceptual integration, we now test our method on its real purpose: recovering the scales from incomplete experimental data (\reffig{noisy}). We thus limit our measures to 150 repetitions for each tested stimulus. Furthermore, we split our population in 5 ensembles of 100 `simultaneously recorded' neurons, so that noise covariance information (\refeq{jpsth}) is only available between neurons belonging to the same ensemble. We use the same candidate values for parameters $w$, $\tR$ and $K$ as before, picking 50 candidate ensembles $\Kap$ for each tested size $K$. Neurons in $\Kap$ always belong to the same `simultaneous ensemble', which is picked randomly. Finally, for each ensemble $\Kap$, we consider 10 additional neurons $i$, from the same `simultaneous ensemble' but segregated from neurons in $\Kap$, to compute the PCV prediction $\West(t)$ (\refeq{West}).

The method then proceeds as above, save a couple of modifications due to the incompleteness of the data. First, concerning SNR computations (\refeq{new-Z}), the estimated covariance matrix $C_\Kap$ may turn out to be rank-deficient up to numerical precision (although it should be full-rank theoretically, since the number of trials (450) is larger than the largest tested size $K$). We thus replace its inverse $C_\Kap^{-1}$ by its Moore-Penrose pseudo-inverse, with the default numerical tolerance of our mathematical software (Matlab). Even so, we observe a global overestimation of predicted sensitivities $Z$, compared to their values in the full-data case (dashed blue lines in \reffig{noisy}a, reproduced from \reffig{noiseless}a). This overfitting is a well-known feature when estimating Fisher sensitivity from insufficient data \citep{Raudys1998,Hoyle2011}. 

Second, concerning mean PCV predictions (\refeq{West}), our final estimates $\West(t)$ become noisy, reflecting the jagginess of the underlying neural measures due to insufficient trials (\reffig{noisy}b). This jagginess is problematic, as it artificially increases measured values for the divergence $\|\West(w,\tR)-\Wstar(w,\tR)\|^2$, which is our final criterion to retrieve plausible values of $(w,\tR)$. However, this effect can be largely compensated by resorting to resampling over trials (bootstrap). More precisely, for each tested parameters $(w,\tR)$, we may describe our noisy measures in the form:
\begin{align*}
\West^{(meas)}(t) &= \West^{(real)}(t) + \eta(t),\\
W^{\star,(meas)}(t) &= W^{\star,(real)}(t) + \eta^\star(t),
\end{align*}
where $\eta(t)$ and $\eta^\star(t)$ describe our (unknown) measurement errors on $\West$ and $\Wstar$. From this follows the estimate:
\begin{equation}
\label{eq:Wboot}
\pE \|\West^{(meas)}-W^{\star,(meas)}\|^2=\|\West^{(real)}-W^{\star,(real)}\|^2+\pE \|\eta\|^2 +\pE \|\eta^\star\|^2,
\end{equation}
where $\pE$ denotes the (theoretical) expectancy over the set of trials giving rise to measures $\West$ and $\Wstar$. This estimate is based on the assumption that measurement errors $\eta(t)$ and $\eta^\star(t)$ are independent, which is likely to be the case given that $\West(t)$ stems from predictions (on the basis of neural tunings and noise covariances) whereas $\Wstar(t)$ stems from measurements of the true PCV curves. All terms involving an expectancy $\pE$ in \refeq{Wboot} can be estimated by resampling with replacement over the set of recorded trials. By computing their difference, we thus get a corrected estimate for $\|\West^{(real)}-W^{\star,(real)}\|^2$. This is the estimate plotted in \reffig{noisy}d. A drawback of this method is that the resulting estimate may become (slightly) negative when the underlying match between $\West$ and $\Wstar$ is ``too good''. However, this does not prevent from estimating the resulting weighting function $\PPCV(w,\tR)$ (\refeq{Ppcv}), accepting that some terms in the exponential may become (slightly) positive.

The final results, in \reffig{noisy}, show that our method is still able to recover the most plausible scales of perceptual integration. Using the same tolerances as previsouly (5\% of the power of the true measures), we find the following final estimates:
\begin{align*}
\widehat{w}&=50 \pm 8 \;\;\textrm{msec},\\
\widehat{t_R}&=81 \pm 6 \;\;\textrm{msec},\\
\widehat{K}&=28.3 \pm 5.2,
\end{align*}
again very close to the true scales of the readout. Notably, $\widehat{K}$ is smaller than its prediction in the full-data analysis: this is a consequence of the slight overfitting on estimated SNRs, which leads to an underestimation of the population size $K$ required to match the psychometric SNR. The issue remains minor in this setup~; however we note that standard cross-validation and regularization techniques exist, that respectively assess and counteract the effects of overfitting \citep{Hastie2009}.

In conclusion, the statistical method introduced above allows to overcome the missing data inherent to realistic recordings, by integrating information from all recorded neurons into a few reliable statistical estimators.

%
%

\section{Discussion}

\subsection{Link with previous literature}
\label{sec:literature}

We have proposed a framework to interpret sensitivity and choice signals in a standard model of perceptual decision-making. The purpose of our study is to help understand how perceptual integration takes place from a full sensory neural population. This question requires, not only to compute neurometric sensitivities or choice signals for individual neurons, but also to integrate these measures in a single big picture of how information is read out from the population as a whole.

The sensitivity to stimulus achievable by a neural population has received much attention, both experimental and theoretical. It was progressively realized that (1) the structure of noise correlations influences the amount of information that can be extracted from a neural population, and (2) the linear readout maximizing sensitivity is generally not a simple average of neural activities, but rather an adequate weighting optimizing the ratio between signal and noise extracted from the population, which corresponds to Fisher's linear discriminant \citep[see][and references therein]{Abbott1999,Averbeck2006}. Similarly, the role of time window $w$ used to integrate the spike counts of each neuron has long been acknowledged to have a direct effect on the overall estimated sensitivity \citep[see, e.g.,][]{Britten1992,Uka2003,Cohen2009,Price2010}.

Choice signals have also received much attention since their first measurements, in the form of choice probabilities \citep{Britten1996}. The temporal evolution of choice signals is routinely computed to qualitatively establish the instants in time when a given population covaries with the animal's percept \citep{DeLafuente2006,Price2010}. Recently, the specific temporal evolution of CP signals during a depth discrimination task has cast doubt on the traditional, feedforward interpretation of CP signals \citep[][see \refsec{hypothesis}]{Nienborg2009}. However, very little studies have {\it quantitatively} interpreted CP signals so far, because no analytical relationship was available to interpret their values. Only recently have \citet{Haefner2013} derived the analytical expression of CPs in the standard model of perceptual integration (see \refsec{CP}).

To the best of our knowledge, only one study has explicitly proposed to jointly use sensitivity and choice signals, as two independent constraints characterizing the underlying neural code. In this seminal study, \citet{Shadlen1996} proposed a feed-forward model of perceptual integration in visual area MT responding to a moving dots stimulus, and studied how the population's sensitivity and individual neuron CPs vary as a function of model parameters such as the number of neurons, strength of noise correlations, etc. In \refsec{readout}, we have formalized this intuition of \citet{Shadlen1996}, by showing that sensitivity and choice signals are two distinct, constitutive elements of the joint covariance structure between percept $\fhat$ and neural activity $\bs(t)$ (\reffig{model}c). The third constitutive element is the noise covariance structure of $\bs(t)$ itself, a result also intuited by \citet{Shadlen1996} even though they assumed an oversimplified, homogeneous noise correlation matrix.

Unlike most previous theoretical studies on the subject, we explicitly modeled all neural activities in time. Indeed, this is the only way of targeting the instants of sensory stimulation which contribute to percept formation, and thus to decipher to $K$--$w$ tradeoff on sensitivity. Finally, the statistical approach developped in \refsec{statistical} is, to our knowledge, the first attempt to build inhomogeneous, partial measures of neural activity into a quantitative interpretation of percept formation from the full neural population.

\subsection{Choice signals in realistic experiments}
\label{sec:CP}

Our model, as presented above, assumes a direct perceptual report of stimulus value $\fstar$ on every trial. Real experiments generally involve a more indirect report: to allow easier task learning by the animal, the report is always binary. In the classic random dot motion discrimination task \citep{Britten1992}, a monkey is visually presented with a set of randomly moving dots whose overall motion is slightly biased towards the left ($f<0$ in our notations) or towards the right ($f>0$). The monkey must then press either of two buttons depending on its judgement of the overall movement direction. In another classic task \citep{Mountcastle1990}, monkeys must discriminate the frequencies $f_1$ and $f_2$ of two successive vibrating stimuli on their fingertip. They must press one button if they consider that $f_1>f_2$, and the other button otherwise.

Thus, classic choice signals such as CP only measure the covariation between the spike train of each neuron and the animal's binary choice $c$ on each trial. To infer anything about the animal's underlying {\it percept} $\fstar$, it is also necessary to assume a behavioral model describing how the monkey takes a binary decision, on every trial, on the basis of its sensory percept. Most often, this behavioral model is implicitly assumed to be optimal. For example, in the random dot motion task, it is generally assumed that $c=H(\fstar)$ (Heavyside function), which is clearly the optimal policy if the animal has no further information about $f$. In the two-frequency task, the optimal behavioral model would be $c=H(\fstar_1-\fstar_2)$. However, in the real experiment, the monkeys have to memorize $f_1$ for a few seconds before $f_2$ is presented, so potential effects of memory loss may also come into play. More generally, behaving animals can display biases, lapses of attention, various exploratory and reward-maximization policies that lead to deviations from the optimal behavioral model. To summarize, choosing a relevant behavioral model is a connex problem that cannot be addressed here, and that will vary depending on the task and individual considered.

\begin{figure}
\centerline{
         \includegraphics[width=\columnwidth]{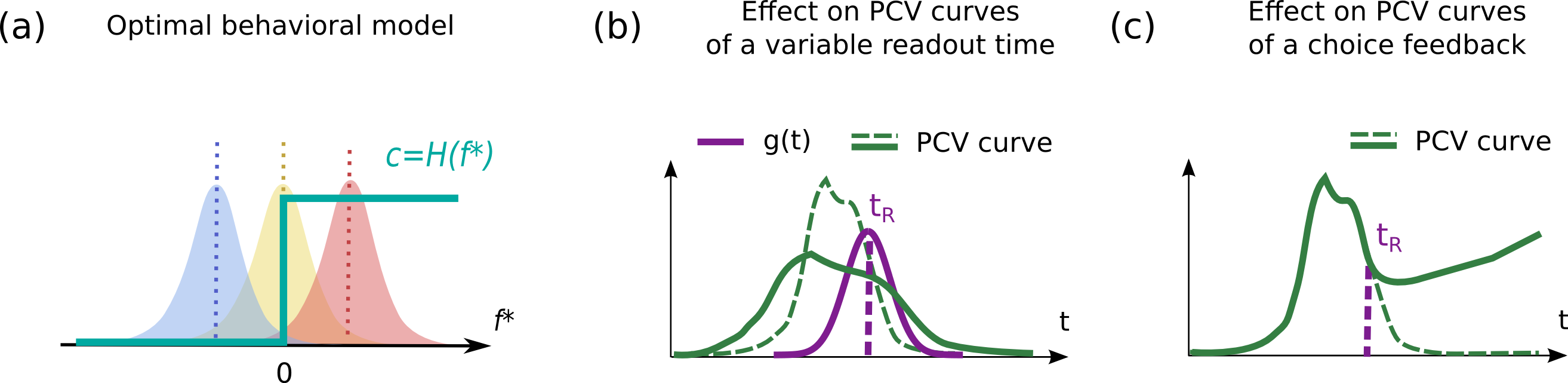}
	}
        \caption{\label{fig:disc} Discussion. (a) Classic behavioral model. If the task is to judge whether $f>0$, the optimal behavioral policy consists of the simple threshold rule $c=H(\fstar)$ (Heavyside function). Furthermore, the trial-to-trial distribution of percept $\fstar$ given $f$ (distributions with different colors) is generally assumed to be Gaussian. Under these hypotheses, sensitivity and PCV signals used in this article are directly computed from real experimental data (neurometric curve and choice signals). (b) If readout time $\tR$ varies strongly from trial to trial (with density $g(t)$), it leads to a flattening of PCV signals (thick green curve) compared to the case with deterministic $\tR$ (dashed green curve). (c) If a decision-related signal feedbacks into sensory areas, it leads to a divergence of PCV signals (thick green curve) after the readout time $\tR$, compared to the case without feedback (dashed green curve).}
\end{figure}

However, for most tractable behavioral models, the predicted sensitivities and choice signals will ultimately rely on the quantities introduced in this article. To take the simplest example, we focus on the random dot motion task with optimal policy $c=H(\fstar)$---as assumed in most models of the task---and make the classic assumption that the statistics of $\fstar$ (given $f$) are Gaussian (\reffig{disc}a). This model predicts the following psychometric curve (probability of button presses as a function of stimulus value):
$$ \pE(c|f)=\pP(\fstar>0|f)=\Phi\big(\sqrt{\Zstar}f\big),$$
where $\Phi$ is the standard cumulative normal distribution, and $\Zstar$ is the square SNR for $\fstar$, as defined in \refeq{Zstar}. Thus, $\Zstar$ used in our model can easily be retrieved from experimental measures of the psychometric curve.

Same results hold for choice signals. Generally, choice signals are directly computed over some temporal average $\overline{s_i}$ of the underlying spike trains. Choice probability for every neuron $i$ measures the area under the ROC curve between the two distributions of $\overline{s_i}$, respectively conditioned on $c=0$ and $c=1$ \citep{Green1966}. Recently \citet{Haefner2013} have shown that, assuming (1) multivariate Gaussian statistics between $\overline{\bf s}$ and $\fstar$, and (2) the optimal behavioral model $c=H(\fstar)$, choice probability can be analytically expressed as:
\begin{equation*}
\mathrm{CP}_i\simeq\frac{1}{2}+\frac{\sqrt{2}}{\pi}\;\frac{\sqrt{\Zstar}\; \overline{\pistar_i}}{\sigma(\overline{s_i})},
\end{equation*}
a formula virtually exact over the full range of plausible CP values. The rightmost fraction is nothing but the Pearson correlation between variables $\overline{s_i}$ and $\fstar$. The numerator involves the linear covariance between $\overline{s_i}$ and $\fstar$ which is, in our notations, the temporally averaged PCV curve $\overline{\pistar_i}$. The authors further derived that, in the standard model of percept formation with readout vector $\ba$, this term is given by $\overline{\bpistar}=\bC \ba$, which is exactly the PCV characteristic equation (\refeq{char-pcov}) in its temporally-averaged form. The CP formula involves a normalization by $\sigma(\overline{s_i})$, the standard deviation of spike count $\overline{s_i}$. This prevents from a straightforward extension of the formula in time, because $\sigma(\overline{s_i})$ tends to infinity as the integration window used to compute $\overline{s_i}$ tends to zero.

A simpler measure of choice signals is the choice-conditioned difference in firing rate \citep{Britten1996}, which can be computed for every individual neuron $i$ as $\Delta_i(t):=\pE(s_i(t)\;|c=1)-\pE(s_i(t)\;|c=0)$. Under the same assumptions as above (Gaussian statistics for ${\bf s}$(t) and $\fstar$, optimal behavioral model), this difference can be analytically expressed\footnote{Relying on the general formula $\pE(X_1|X_2>0)=\sqrt{2\pi^{-1}}\rho$, applicable to any bivariate normal variables $(X_1,X_2)$ with means 0, unitary variances, and correlation coefficient $\rho$. We note that the assumption of normality is violated at small time scales because $s_i(t)$ is clearly not Gaussian in that case. However, in practice, $\Delta_i(t)$ is always computed with a minimal amount of temporal smoothing which resolves this potential issue.} as:
\begin{equation}
\label{eq:Delta}
\Delta_i(t)=\sqrt{\frac{2}{\pi}} \sqrt{\Zstar}\; \pistar_i(t).
\end{equation}
This is very close to the CP formula, but without the additional normalization by $\sigma(\overline{s_i})$. Thus it directly allows for a simple generalization to temporal signals. Since $\Delta_i(t)$ is easily computable from experimental data, it provides the easiest way of accessing the underyling PCV curves $\pistar_i(t)$ used in our article.

\subsection{Model hypothesis}
\label{sec:hypothesis}

\subsubsection{Linear integration}

The readout model (\refeq{readout}) used to analyze sensitivity and choice signals is an instalment of the `standard', feedforward model of percept formation. As such it makes a number of hypothesis which should be understood when applying our methods to real experimental data. First, it assumes that the percept $\fstar$ is built linearly from the activities of the neurons. There is no guarantee that this is the case during real percept formation, but linearity is an unavoidable ingredient to make quantitative predictions at the population level. Even if the real percept formation departs from linearity, fitting a linear model will most likely retain meaningful estimates for the coarse information (temporal scales, number of neurons involved) that we seek to estimate in this work.

More precisely, the model in \refeq{readout} assumes that spikes are integrated using a kernel separable across neurons and time, that is $A^i(t)=a^i h_w(t)$. Theory does not prevent from studying a more general integration, where each neuron $i$ contributes with a different time course $A^i(t)$. The readout's characteristic equations are derived equally well in that case. Rather, assuming a separable form 
reflects (1) the intuition that the temporal components of integration are rather uniform across the population, and (2) the impossibility to fit a model with general kernel $A^i(t)$. Instead, we summarize temporal integration from the population by two parameters $w$ and $\tR$, opening the door to a reliable estimation from data. Although the integration shape $h$ could also be fit from data in theory, it seems more fruitful to assume a simple shape from the start (a classic square window kernel in our applications). Given that our goal is to estimate the coarse scales of percept formation, our method will likely be robust to various simple choices for $h$. 
As a simple example, we tested our method, assuming a square window kernel, on data produced by a readout using an exponential kernel, and still recovered the correct parameters $w$, $\tR$ and $K$.

\subsubsection{Non-deterministic extraction time}
\label{sec:withg}

Our model, as presented above, makes another important assumption: that perceptual readout occurs at the same time $\tR$ on every stimulus presentation.
This assumption is likely to be valid in perceptual tasks that allow a fast reaction from the animal (`reaction time' tasks), in which case $\tR$ will generally be as small as it can get \citep[see, e.g.,][]{Stanford2010}. However, when sensory stimulation lasts longer (say, over 500 msec) it opens the door to variations in $\tR$ from trial to trial, or even to several reactualizations of percept $\fstar$ during the same trial.
For example, imagine that the stimulus is a particular RGB color on a monitor, and you are asked to judge whether it contains more green (G) or blue (B). From intuition, we can tell that our performance in such a task will not sensibly increase whether we watch the color for one second or one minute. In our model's formalism (\refeq{readout}), this reveals built-in limitations on the effective integration window $w$ that we can use in the task (remember that the readout's performance is proportional to $w$). But then, if our percept arises from a limited integration window $w$ and we indeed watch the color for a full minute, when is our percept built?

In appendix \ref{sec:appChar}, we derive a more general version of the characteristic equations (\refeq{char-tuning}-\ref{eq:char-pcov}) assuming that $\tR$ in \refeq{readout} is itself a random variable, drawn on each trial following some probability distribution $g(t)$. Because sensory neurons have rather stationary activities in time, this additional assumption does not strongly affect the readout's sensitivity. On the other hand, it affects strongly PCV curves. Essentially, the resulting PCV curve resembles a convolution of the deterministic PCV curve by $g(t)$  (\reffig{disc}b, \refsec{appg}). If $g(t)$ is substantially distributed in time, the PCV curves will become broader, and flatter. In practice, this means that if a behavioral task is built such that $\tR$ can display strong variations from trial to trial, the statistical method introduced above will produce biased estimates. In theory, this issue could be resolved by adding an additional parameter in the analysis to describe $g(t)$ (see \refsec{appg}), but the validation remains to be done.

\subsubsection{Top-down influence of choice}

Finally, our `standard' model assumes that percept formation is exclusively feed-forward. The activities $s_i(t)$ of the sensory neurons are integrated to give rise to percept $\fhat$ and the animal's resulting choice $c$, and this forming decision does not affect sensory neurons in return. Recent evidence suggests that reality is more complex. By looking at the temporal evolution of CP signals in V2 neurons during a depth discrimination task, \citet{Nienborg2009} evidenced dynamics which are best explained by a top-down signal, biasing the activity of the neurons on each trial {\it after the choice is formed}. In our notations, the population spikes $s_i(t)$ would thus display a choice-dependent signal which kicks in on every trial after time $\tR$, resulting in PCV signals that deviate from their prediction in the absence of feedback (\reffig{disc}c).

What descriptive power does our model retain, if such top-down effects are strong? The answer depends on the nature of the putative feedback. If the feedback depends linearly on percept $\fhat$ (and thus, on the spike trains), its effects are fully encompassed in our model. Indeed, this feedback signal will then be totally captured by the neurons' linear covariance structure $\gamma_{ij}(t,s)$, so that our predictions will naturally take it into account. This is also the case if the oddity noted by \citet{Nienborg2009} is due to global shifts of neural excitability from trial to trial. On the other hand, if the feedback depends directly on the choice $c$---which displays a nonlinear, `all-or-none' dependency on $\fhat$---then it will not be captured by our model, and lead to possible biases. Even so, the effects of the feedback could be largely alleviated through a small trick: compare true and predicted PCV signals only up to (candidate) time $\tR$ (see \refeq{L2}). 

\subsection{Extrapolation to larger neural ensembles}
\label{sec:extrapolation}

Can we understand in more depth the statistical principles at work underneath our method of estimation? What factors govern the evolution of sensitivity $Z(\Kap)$ (\reffig{noiseless}a, \refeq{new-Z}) and mean PCV signal $W(t|\Kap)$ (\reffig{noiseless}b, \refeq{W-kap}), as a function of the number of neurons $K$ used for readout? This question is not only of theoretical, but also of practical interest. Indeed, it may happen in real applications that the number of simultaneously recorded neurons $K_\mathrm{max}$ is too small to observe the crossing of predicted and true SNR curves (\reffig{noiseless}a)\footnote{Actually, this is always bound to happen for small tested parameter $w$, following the $K$--$w$ tradeoff.}. In such a case, predictions will be biased because no recorded ensemble $\Kap$ can readily account for animal sensitivity.

What predictive power do ensembles up to size $K_\mathrm{max}$ contain about larger ensembles? For example, can we extrapolate the shape of the mean SNR curve (\reffig{noiseless}a) to $K>K_\mathrm{max}$? In appendix \ref{sec:appSVD} we address this question theoretically, by studying the value of SNR and PCV signals as a function of ensemble size $K$, and of the general structure of activity in the population. Our study relies on the singular value decomposition (SVD) of neural activity in the population. The SVD reveals a set of $m=1\dots M$ independent {\it modes} of population activity, each mode being associated to a power $\lambda^2_m$ and a sensitivity $\eta^2_m$. Essentially, the sensitivity embedded in a neural ensemble $\Kap$ of size $K$ increases as the sum of sensitivities for the $K$ first modes in the population---which are the modes with the largest powers $\lambda_m$. Conversely, the overall power of PCV signal $W(t)$ decreases as the average value of $\lambda^2_m$ in these $K$ first modes, weighted by their respective sensitivities. 

Because there is no general relationship between the power $\lambda_m$ of a mode and its sensitivity to stimulus $\eta_m$, there is no trivial way of extrapolating SNR and PCV predictions to ensemble sizes $K$ that were not monitored simultaneously. Any such extrapolation can only be done through specific assumptions about the link between $\lambda_m$ and $\eta_m$---which essentially amounts to characterizing the relative embedding of signal and noise in the full population \citep{Wohrer2012}. For example, it is classically assumed that the noise covariance matrix is ``smooth'' with respect to the signal covariance matrix, so that the former can be predicted on the basis of the latter \citep{Wohrer2010,Haefner2013}. Thus, while extrapolation of the statistical method above to larger populations is not trivial, it can be performed under specific assumptions about the embedding of signal and noise in the population considered.

\section{Conclusion}

We have shown how classic data recorded during perceptual decision-making experiments can be interpreted as samples from the joint covariance structure of neural activities and animal decision. Assuming a standard linear model of percept formation from neural activities, we derived a set of characteristic equations which relate neural and perceptual data, and thus define implicitly the parameters of perceptual integration by the animal on the basis of its sensory neurons. The neural data consist of neural PSTHs (first moment of neural activities) and JPSTHs (second moment of neural activities). The perceptual data consist of the animal's sensitivity, and of each neuron's covariance with the animal's choice---a quantity often assessed through choice probabilities, and for which we proposed a simpler linear equivalent coined {\it percept covariance} (PCV).

We then proposed a method to utilize these characteristic equations in a case of practical interest, when experimenters only have access to finite statistical samples of neural data across the full population. Our goal was to successfully recover the instants in time and the typical number of neurons being used for percept formation---a difficult problem which cannot be solved on the sole basis of sensitivity information, due to the ``$\nK$--$w$ tradeoff''. Our method relies on statistical averages of predicted sensitivity and PCV signals arising from random, candidate neural ensembles used as the source of percept formation~; and seeks to match these predictions with the true, recorded perceptual data. We tested this method on an artificial neural network producing a form of stimulus encoding, and showed that it successfully recovers the scales of perceptual integration, on the basis of sample recordings of realistic size.

Our method opens the way to novel experimental assessments of percept formation in sensory decision-making tasks. Indeed, the two main quantities used in our statistical analysis---sensitivity $Z$ (\refeq{new-Z}) and mean PCV curve $W(t)$ (\refeq{W-kap})---rely on classic experimental measures. The main limitation of our approach is the size of candidate readout ensembles which can be considered, as they should necessarily have been recorded simultaneously. However, the number of simultaneously recorded neurons is constantly pushing upwards with modern experimental techniques, so we may expect that this limitation, if it exists, will soon be overcome. Furthermore, through a theoretical analysis based on the singular value decomposition (SVD) of neural activities, we showed the possibility of extrapolation to larger ensemble sizes than those simultaneously recorded, although such extrapolations can only be done under specific assumptions, and on a case-by-case basis. For all these reasons, our method can readily be tested on real data, and hopefully provide new insights into the nature of percept formation from populations of sensory neurons.



\newpage

\begin{appendices}
\numberwithin{equation}{section}

%
%

\section{Characteristic equations for the readout}
\label{sec:appChar}

\subsection{Derivation}

We here derive the characteristic equations for the linear readout introduced in the main text, and further comment some of its properties. We consider a more general version of \refeq{readout}, where the extraction time $\tR$ is allowed to vary from trial to trial. We thus assume that $\tR$ is itself a random variable, drawn on each trial according to some density function $g(t)$, independently of neural activities $\bs (t)$. The full readout model then writes:

\begin{align}
\fhat(\tR) &= \sum_i \int_{u>0} a^i s_i(\tR-u) h_w(u) du, \label{eq:gen-readout}\\
\tR &\sim g(t). \label{eq:gen-g}
\end{align}
This model naturally encompasses the simpler version presented in the main text, with a deterministic time $\tR$: it corresponds to taking $g(t)$ as a Dirac function located on that deterministic value.

The characteristic equations for this model rely on a straightforward computation of the second order statistics of $\fhat$, starting from \refeq{gen-readout}. To deal with random time $\tR$, we note that for any random process $X(t)$ independent of $\tR$, $\pE(X(\tR))=\int_{t=-\infty}^{+\infty}g(t)\pE(X(t))dt$. This expression is valid only if $\tR$ is independent from the random variables contributing to $X$ (in our case, the spike trains).

Then, the expected value of $\fhat$ given a stimulus $f$ writes:
\begin{align}
\pE(\fhat(\tR)|f)&=\int_t g(t)\sum_i \int_{u>0} a^i \pE(s_i(t-u)|f) h_w(u)\,du\,dt \nonumber\\
&=\sum_i a^i \int_t (g\star h_w)(t) \lambda_i(t;f)dt, \label{eq:gen-Efhat}
\end{align}
where $g\star h_w (t)=\int_u g(u)h_w(u-t)du$ is the temporal correlation between $g$ and $h_w$, and $\lambda_i(t;f)$ is the PSTH for neuron $i$ in stimulus condition $f$, defined as in the main text (\refeq{psth}).

Similarly, the expected value of $\fhat^2$ given a stimulus $f$ writes:
\begin{align}
\pE(\fhat(\tR)^2|f)&=\int_t g(t)\pE\Big(\Big(\sum_i \int_{u>0} a^i s_i(t-u) h_w(u) du\Big)^2\Big|f\Big)\,dt \nonumber\\
&=\int_t g(t)\sum_{ij} \iint_{(u,v)>0} a^i a^j \pE(s_i(t-u)s_j(t-v)|f) h_w(u)h_w(v) \,du \,dv \,dt \nonumber\\
&=\sum_{ij} a^i a^j \iint_{(t,s)} G_w(t,s) \eta_{ij}(t,s;f)\,dt\,ds, \label{eq:gen-Efhat2}
\end{align}
where we have defined $G_w(t,s):=\int_u g(u)h_w(u-t)h_w(u-s)du$, and $\eta_{ij}(t,s;f):=\pE(s_i(t)s_j(s)|f)$. $\eta_{ij}(t,s;f)$ is very related to the covariance structure in the population. It corresponds to the ``plain'' JPSTH for the neurons in stimulus condition $f$, before correcting by the so-called ``product predictor'' \citep{Aertsen1989}.

Finally, the expected value for the product of $\fhat$ and the activity of any neuron $s_i(t)$ writes:
\begin{align}
\pE(\fhat(\tR)s_i(t)|f)&=\int_s g(s)\sum_j \int_{u>0} a^j \pE(s_i(t)s_j(s-u)|f) h_w(u) du\,ds \nonumber\\
&=\sum_j a^j \int_s (g \star h_w)(s) \eta_{ij}(t,s;f)\,ds, \label{eq:gen-Efhatsj}
\end{align}
using the same notations as above.\\

The three expressions \refeq{gen-Efhat}-\ref{eq:gen-Efhatsj} roughly correspond to the three characteristic equations for the readout. To obtain them, we consider the variational versions of the previous expressions. First, we obtain the characteristic equation for tuning by differentiating \refeq{gen-Efhat} with respect to stimulus. Second, equations \ref{eq:gen-Efhat2} and \ref{eq:gen-Efhatsj} are expressed in `product' form $\pE(XY)$, whereas the corresponding characteristic equations are expressed in `covariance' form $\Cov(X,Y)=\pE(XY)-\pE(X)\pE(Y)$.
Once this is done, and after some rearrangement of the terms, we obtain the characterisitic equations for tuning (\refeq{gen-tuning}), sensitivity (\refeq{gen-Z}) and percept covariance (\refeq{gen-pcov}):
\begin{align}
\partial_f \pE(\fhat|f) &= \sum_i a^i \int_t (g\star h_w)(t) \beta_i(t)dt, \label{eq:gen-tuning}\\
\langle\Var(\fhat|f)\rangle_f &= \sum_{ij} a^i a^j \Big( \iint_{(t,s)} G_w(t,s)\gamma_{ij}(t,s)\,dt\,ds + V_{ij}^{temp} \Big),\label{eq:gen-Z}\\
\langle\Cov(\fhat,s_i(t)|f)\rangle_f &=\sum_j a^j \int_s (g \star h_w)(s) \gamma_{ij}(t,s)\,ds.\label{eq:gen-pcov}
\end{align}
In \refeq{gen-tuning}, $\beta_i(t)$ is the temporal tuning curve for neuron $i$, defined as in \refeq{tuning}. If the readout is unbiased, the left-hand side is equal to 1, as in the main text. In \refeq{gen-Z} and \ref{eq:gen-pcov}, $\gamma_{ij}(t,s)$ is the covariance structure (JPSTH) between neurons $i$ and $j$, defined as in \refeq{jpsth}.
 
Finally in \refeq{gen-Z}, matrix $V_{ij}^{temp}$  is an additional source of variance that appears only when $g(t)$ has an extended temporal support, i.e., when $\tR$ is non-deterministic. It then writes:
\begin{equation*}
V_{ij}^{temp} = \iint_{(t,s)} G_w(t,s) \Big\langle \Big(\lambda_i(t;f)-\overline{\lambda_i}(f)\Big)\Big(\lambda_j(s;f)-\overline{\lambda_j}(f)\Big) \Big\rangle_f \,dt\,ds,
\end{equation*}
where $\overline{\lambda_i}(f):=\int_t (g\star h_w)(t) \lambda_i(t;f)dt$ is the temporal average already used above (\refeq{gen-Efhat}). Thus, $V_{ij}^{temp}$ measures a form of temporal covariance in the PSTHs for the neurons.

When $\tR$ is deterministic, as in the main text, we have $g(t)=\delta(t-\tR)$, a Dirac function. Then, the temporal integration kernels used in \refeq{gen-tuning}-\ref{eq:gen-pcov} boil down to $(g \star h_w)(t)=h_w(\tR-t)$ and  $G_w(t,s)=h_w(\tR-t)h_w(\tR-s)$. One checks easily that in these conditions, the additional temporal variance term $V_{ij}^{temp}$ vanishes, and we recover the characteristic equations from the main text.

\subsection{Additional interpretations}

\subsubsection{Sensitivity as a function of $w$}
\label{sec:appw}

In the form of \refeq{gen-Z}, it is not clear how the value of $w$ influences the variance of $\fhat$, and thus the readout's sensitivity. To get a better intuition, let us first neglect the temporal variance term $V_{ij}^{temp}$. One checks easily that kernel $G_w(t,s)$, introduced above, verifies the following property: $\int_t G_w(t,t+\tau)dt=(h_w\star h_w)(\tau)$, the autocorrelation of kernel $h_w$. As a result, we can rewrite \ref{eq:gen-Z} in the form :
\begin{align}
\langle\Var(\fhat|f)\rangle_f &= \sum_{ij} a^i a^j \int_\tau (h_w\star h_w)(\tau) \Big(\int_t \frac{G_w(t,t+\tau)}{h_w\star h_w(\tau)}\gamma_{ij}(t,t+\tau)dt\Big)d\tau\nonumber\\
&= \sum_{ij}a^i a^j \int_\tau (h_w\star h_w) (\tau) \overline{\gamma_{ij}}(\tau)d\tau.\label{eq:gen-ccgm}
\end{align}
In the first line, the function of $t$ defined by the fraction is positive and has an integral of 1, so it operates as a temporal averaging on $\gamma_{ij}(t,t+\tau)$. The resulting average over $t$, noted $\overline{\gamma_{ij}}(\tau)$ in the second line, is thus a form of {\it cross-correlogram} between neurons $i$ and $j$, measuring the average covariance between the spikes from $i$ and $j$ separated by a time lag $\tau$.

Because $h_w$ is a low-pass kernel with scale $w$, its autocorrelation function typically has support on $[-w,w]$, and verifies\footnote{Assuming proper scaling for shape function $h$: $\int_u h(u)du=\int_u h(u)^2du=1$.}: $(h_w\star h_w)(0)=w^{-1}$. On the other hand, $\overline{\gamma_{ij}}(\tau)$ typically has support on some interval $[-\tau_\gamma,\tau_\gamma]$, where $\tau_\gamma$ is the typical time scale of noise correlations in the population. As a result, as soon as $w$ gets bigger than $\tau_\gamma$, the integral in (\ref{eq:gen-ccgm}) starts behaving like $w^{-1}$, and the SNR of $\fhat$ scales as $w$. A similar analysis can be performed on the additional term $V_{ij}^{temp}$ (\refeq{gen-Z}).

\subsubsection{Non-deterministic $\tR$}
\label{sec:appg}

What are the main departures from the main text when function $g(t)$ has an extended temporal support? From \refeq{gen-tuning}-\ref{eq:gen-pcov}, it is clear that the general form of the characteristic equations still holds:
\begin{align*}
1 &= \bb^\top\ba,\\
Z^{-1} &= \ba^\top\bC\ba,\\
\bpi(t)&= \bGam(t)\ba,
\end{align*}
but with more general definitions of $\bb(w,g)$, $\bC(w,g)$ and $\bGam(t|w,g)$. First, an additional covariance matrix $\mathrm{\bf V}^{temp}$ may contribute to $\bC$, if neural activities are not stationary in time. Indeed, if $\tR$ varies from trial to trial, any variation of firing rates in time creates an additional source of variability in $\fhat$. 

Second, through \refeq{gen-pcov}, $g(s)$ acts a weighting factor over the PCV curves that would be obtained for each $\tR$: $\bGam(t|g)=\int_sg(s)\bGam(t|\tR=s)ds$. This leads to the spreading of PCV curves sketched in \reffig{disc}c.

These two features lead to lose one specific property of the deterministic case. When the ``natural'' temporal averaging of PCV signals was considered, that is $\overline{\bpi}=\int_t h_w(\tR-t)\bpi(t)dt$, the integrated PCV equation yielded $\overline{\bpi}=\bC \ba$, because $\overline{\bGam}=\bC$. In the general case, the ``natural'' temporal averaging is $\overline{\bpi}=\int_t (g\star h_w)(t)\bpi(t)dt$, and one checks easily that $\overline{\bGam}\neq\bC$. Thus, with general $g(t)$, the sensitivity (\refeq{gen-Z}) and PCV (\refeq{gen-pcov}) equations become more dissociated.\\

In these conditions, it is unclear whether the statistical approach introduced in the main text could be extended, to also recover a non-deterministic extraction function $g(t)$. The main concern is that the temporal evolution of PCV signals is only determined by the aggregate function $(g\star h_w)(t)$ (\refeq{gen-pcov}), which cannot be used to disentangle $g(t)$ and $w$ separately. However, general considerations suggest that the method could still work in that case. Indeed, the respective effects of $g(t)$ and $w$ on the covariance structures used in \refeq{gen-Z}-\ref{eq:gen-pcov} can roughly be thought of as a scaling: 
\begin{equation}
\label{eq:app-scaling}
\overline{\bGam}\simeq\epsilon(w,g)\bC,
\end{equation}
because the overall ``shape'' of covariance between neurons (as opposed to its ``strength'') does not depend much on the precise temporal integration used to compute their activity. Actually, under the specific assumption that $\gamma_{ij}(t,s)=\overline{\gamma_{ij}}Q(|t-s|)$ (stationary activities with uniform temporal correlations), relationship (\refeq{app-scaling}) can be shown to be exact, with 
$$
\epsilon(w,g)=\frac{\int_\xi\widetilde{Q}(\xi)\|\widetilde{h_w}(\xi)\|^2 \|\widetilde{g}(\xi)\|^2d\xi}
{\int_\xi \widetilde{Q}(\xi)\|\widetilde{h_w}(\xi)\|^2d\xi}
$$
expressed in terms of Fourier transforms. As a result, the mean PCV curve $W(t)$ (\refeq{new-pcov}, \ref{eq:W-kap}) is predicted to scale as $\epsilon(w,g)$. So, while matching the temporal support of $W(t)$ and $\Wstar(t)$ constrains the value of $(g \star h_w)(t)$, matching their overall power constrains $\epsilon(w,g)$, and we can hope to disentangle the values of $g(t)$ and $w$ separately. In practice though, this would require the fitting of at least one additional temporal parameter~; typically, the standard deviation of $\tR$ from trial to trial.


%
%

\section{Singular value analysis---summary}
\label{sec:appSVD}

We summarize here the main results of a theoretical analysis to understand the evolution of SNR and PCV signals achieved by readout ensembles of growing size $\nK$. Detailed mathematical derivations are available in Supplementary Section S2. 
For simplicity we focus only on time-integrated neural activities $\overline{s_i}:=\int_u h_w(u)s_i(\tR-u)du$, assuming a fixed choice of $(w,\tR)$. We consider random readout ensembles $\Kap$ in the population, and two resulting indicators. First, we consider the sensitivity $Y(\Kap)$, linked to SNR $Z$ by relationship $Y=Z(1+Z)^{-1}$. This is the natural description of sensitivity in the framework below. It is obtained like $Z$ in the main text (\refeq{new-Z}) but replacing the noise covariance matrix $\bC$ by the total covariance matrix $\bA=\bC+\langle f^2\rangle \bb \bb^\top$. Second, we consider the mean PCV in the population $\Wbar(\Kap)$, obtained as the ``natural'' temporal integration of signal $W(t|\Kap)$ from the main text (\refeq{W-kap}): $\Wbar:=\int_u h_w(u)W(\tR-u)du$. Since $W(t)$ is mostly positive, $\Wbar$ roughly corresponds to the overall power in $W(t)$.

\paragraph{SVD reformulation of neural activity.} 
The analysis relies on the singular value decomposition (SVD) of population activity into $m=1\dots M$ orthogonal {\it modes}:
$$
\overline{s}_i^{f\om}= \sum_{m=1}^M \lambda_m u_i^m v_m^{f\om},
$$
where the lower index $i=1\dots\Ntot$ indicates neurons in the population, and the upper index indicates all possible stimuli $f$ and random realizations $\om$ of network activity. Each mode $m$ is defined by its power $\lambda_m>0$, its distribution vector (over neurons) $\bu^m$, and its appearance variable $v_m$ which takes a different random value on every trial. By construction, the various modes are orthogonal across neurons ($(\bu^m)^\top\bu^n=\delta^{mn}$), and linearly independent across trials ($\Cov_{f\om}(v_m,v_n)=\delta_{mn}$), so they typically correspond to distinct ``patterns of activity'' in the population. The power $\lambda_m$ describes the overall impact of mode $m$ on population activity. We assume $\lambda_1\geq\dots\geq\lambda_M$, so we progressively include modes with lower power---either because they involve only a small fraction of neurons, either because they appear only on rare trials. The number of modes $M$ is the intrinsic dimensionality of the neural population's activity. In real populations we expect $M<<\Ntot$, because neural activities are largely correlated.

The SVD is best viewed as a change of variables reexpressing neural activities $\{\overline{s_i}\}_{i=1\dots \Ntot}$ in terms of mode appearance variables $\{v_m\}_{m=1\dots M}$. Just like individual neurons, each mode $m$ can be associated to a {\it sensitivity} to stimulus $\eta_m$, which describes the proportion of the mode's power $\lambda_m$ due to variations of the signal ($f$), as opposed to variations of the noise ($\om$). Since modes are linearly independent, the full population's sensitivity corresponds to the sum of individual mode sensitivities: $Y(\infty)=\sum_m \eta_m^2$. 

\paragraph{Sensitivity and PCV from finite neural ensembles.}
We now want to estimate the amount of stimulus sensitivity $Y(\Kap)$ that can be extrated, not from the full population, but from neural subensembles of size $\Kap$. The SVD provides a natural reinterpretation of this problem in terms of activity modes: each ensemble $\Kap$ ``reveals'' only a fraction of the underlying modes.
The pivotal object to perform this reinterpretation is our so-called {\it data matrix}:
$$
\bD_\Kap := \Big\{ d_i^m = \lambda_m u^m_i\Big\}_{i \in \Kap}^{m=1\dots M},
$$
an $M \times K$ matrix describing the activity of neural ensemble $\Kap$ in the space of modes. In the original problem formulation, the $K \times K$ matrix $\bD^\top \bD$ describes the covariance of neural activity in ensemble $\Kap$, and we want to estimate the resulting sensitivity. In the dual formulation, the $M \times M$ matrix $\bD \bD^\top$ describes a covariance structure between modes, but estimated only from the sample neurons in $\Kap$. The problem now lives in a space of fixed dimensionality $M$, and can be related to classical problems of estimating covariance structures from a finite number of samples---in our case, the neurons.

Applying this dual approach, we find that $Y(\Kap)$ and $\Wbar(\Kap)$ depend on readout ensemble $\Kap$ only through an $M \times M$ matrix $\bDelta_\Kap$, the (rank $K$) orthogonal projector on the span of vectors $\{\bd_i\}_{i \in \Kap}$ in mode space:
\begin{align*}
Y(\Kap) &= \boeta^\top \bDelta_\Kap \boeta,\\
\Wbar(\Kap) &= -B^2 + (\Ntot Y(\Kap))^{-1} \boeta^\top \bLam^2 \bDelta_\Kap \boeta,
\end{align*}
where $B^2:=\pE_i(b_i^2)$ is the average square tuning in the population.
Furthermore, the average projector $\bDelta_\Kap$ across ensembles of size $K$, that is $\pE_K\bDelta$, is approximately diagonal in mode space. Noting $\{\epsilon_K^m\}$ for its diagonal, we thus obtain the approximations:
\begin{align}
\pE_K Y&\simeq \sum_{m=1}^M \epsilon_K^m \eta_m^2,\label{eq:EKY-summ}\\
\pE_K \Wbar &\simeq -B^2 + \Ntot^{-1} \frac{\sum_{m=1}^M \epsilon_K^m \eta_m^2 \lambda_m^2}{\sum_{m=1}^M \epsilon_K^m \eta_m^2},\label{eq:EKW-summ}
\end{align}
where $0\leq \epsilon_K^m\leq 1$ is the average ``proportion'' of mode $m$ revealed by $K$ random neurons. As modes with larger power $\lambda_m$ tend to be revealed first, a rough but useful image is to consider that $\epsilon_K^m \simeq \mathbb{1}_{m\leq K}$---only the $K$ first modes are revealed by ensembles of $K$ neurons.

Thus, sensitivity $\pE_K Y$ grows with $K$ as mode sensitivities $\eta_m$ are progressively revealed. Saturation occurs when all nonzero $\eta_m$ are revealed, in which case $\pE_K Y = Y(\infty)$. Conversely, the mean PCV $\pE_K \Wbar$ decreases with $K$. Indeed, the fraction in \refeq{EKW-summ} can be viewed as an average power $\langle\lambda^2\rangle_{m,K}$, where each mode $m$ contributes with a weight $\epsilon_K^m \eta_m^2$. As $\epsilon_K^m$ progressively reveals modes with lower power $\lambda_m$, this average power is expected to decrease with $K$. Again, saturation occurs when all nonzero $\eta_m$ are revealed, and then $\pE_K \Wbar = Z(\infty)^{-1}B^2$, the predicted value for choice signals in case of optimal readout from the full population \citep{Haefner2013}.

\paragraph{Extrapolation to large $K$.}
What do these results tell us about possible extrapolations to ensemble sizes $K$ larger than the maximum number of neurons simultaneously recorded by the experimenter? Essentially, that such extrapolations always require further assumptions about the structure of activity in the population.

Indeed, one can imagine scenarios in which the most sensitive modes (those with highest $\eta_m^2$) are associated to relatively low powers $\lambda^2_m$ and thus, appear only at large $K$. This could be the case, for example, if a very local circuit of neurons carries a lot of information about the stimulus, independently from the rest of the population. Because it involves few neurons, the corresponding mode of activity will have a low power $\lambda^2_m$, and will require very large ensembles $\Kap$ to be detected---simply because the corresponding neurons are not recorded in smaller ensembles. A similar discussion can be found in \citet{Haefner2013}. Another example is the encoding network theoretically proposed by \citet{Boerlin2011}, where each neuron spikes only if its information is not already encoded in the activity of the remaining neurons. This results in the appearance of a few, global modes of activity\footnote{Typically the sum of all neural activities, in the simplest instantiation of the model.} which are specifically designed to have a very large SNR, meaning high $\eta_m$ and low $\lambda_m$. In this case, any estimation of sensitivity from a subpopulation $\Kap$ will consistently be smaller than the full population's sensitivity.

To summarize, extrapolation can only be performed under additional assumptions about the overall link between $\eta_m$ and $\lambda_m$---or equivalently, about the relationship between `signal' and `noise' contributions to population activity \citep[see also discussions in][]{Wohrer2012,Haefner2013}. The extent to which such assumptions are justified will depend on each specific context.

\end{appendices}


\small

\normalsize


\newpage

\setcounter{section}{0}
\setcounter{footnote}{0}
\setcounter{equation}{0}

\renewcommand{\theequation}{\arabic{equation}}
\renewcommand{\thesection}{S\arabic{section}}
\renewcommand{\thesubsection}{\arabic{subsection}}
\renewcommand{\thesubsubsection}{\arabic{subsection}.\arabic{subsubsection}}

\begin{flushleft}
{\LARGE
\textbf{SUPPLEMENTARY INFORMATION}
}
\end{flushleft}


\section{Encoding network}
\label{sec:appNetwork}

We detail here the architecture of the artificial encoding network used to test our method (summarized in section 3.1 from the main text). 
This ad-hoc network was designed to display some classic features of sensory cortical neurons involved in perceptual decision-making tasks (e.g, V2, MT, S1, S2\dots). To reproduce the diversity of response naturally observed at the population level \citep{Wohrer2012}, neurons in our network have broadly distributed firing rates, and some diversity in their temporal response profiles. We also wished to reproduce the continuum of tuning to stimulus observed in real populations, where some neurons have positive tuning to stimulus (rate increase when $f$ increases), and other neurons have negative tuning. Finally, we wished to reproduce realistic strengths of noise correlations between neurons in the population (Figure 3b from the main text), 
and insure that the tunings of each pair of neurons (their ``signal'' correlation) be only slightly predictive of their noise correlation---another feature often observed in real sensory populations \citep{Wohrer2012}.


The network consists of two distinct layers of spiking neurons, of which only the second layer (encoding layer) is ``visible'' to the experimenter. The first layer (L1) consists of $100=2\times 50$ independent Poisson neurons, whose firing intensity $f$ constitutes the stimulus encoded by the second layer. On each trial, $f$ takes one of three possible values $25$, $30$ and $35$ Hz. All neurons are equivalent, but segregated in two distinct populations according to their projections on the second layer. The Poisson firing constitutes the only source of randomness in the network from trial to trial.

The second layer (L2) consists of 500 leaky integrate-and-fire (LIF) neurons, some of which receive input from L1, and who are all coupled through a sparse, balanced connectivity. The generic equation for these neurons writes
$$
\tau \frac{dV_i^{(s)}}{dt}=\sum_{j\in \mathrm{L1}} W^{(1,s)}_{ji}\delta(t-t_j)+\sum_{k\in \mathrm{L2}} W^{(2)}_{ki}\delta(t-t_k-\Delta_{ki})+I^{(s)}-(V_i^{(s)}(t)-V^{rest}).
$$
The neuron emits a spike at each time $t_i$ when $V_i^{(s)}$ reaches threshold $V^{thr}$, after what the neuron's potential is reinitialized at resting value $V^{rest}$. All neurons share the same membrane time constant $\tau=20$ msec, threshold $V^{thr}=-50$ mV, and resting potential $V^{rest}=-60$ mV. Upper index $s$ denotes one of three possible subtypes of neurons in L2: Positively-biased neurons ($s=p$, 100 neurons), negatively-biased neurons ($s=n$, 100 neurons) and unbiased neurons ($s=u$, 300 neurons).

Positively-biased neurons receive sparse excitatory connections from 50 neurons in L1 ($W^{(1,p)}_{ji}\geq0$), whereas negatively-biased neurons receive sparse inhibitory connections from the 50 other neurons in L1 ($W^{(1,n)}_{ji}\leq0$). Unbiased neurons receive no direct input from L1 ($W^{(1,u)}_{ji}=0$). As these asymmetries create biases in the total synaptic inputs to each type of cell, the intrinsic currents $I^{(p)}$, $I^{(n)}$ and $I^{(u)}$ also vary depending on neuron subtype, to insure homogeneous firing properties inside the three populations (see Table \ref{tab:L2params}). Finally, all L2 neurons are connected through a single matrix $\mathbf{W}^{(2)}$ of recurrent connections---independently of their subtype. All connection matrices $\mathbf{W}^{(1,s)}$ and $\mathbf{W}^{(2)}$ are sparse with (Erd\"os-Renyi) connection probability $p=0.2$. Non-zero connection strengths are picked uniformly in an interval $[w_\mathrm{min},w_\mathrm{max}]$, which depends on the connection considered: see Table \ref{tab:L2params}. Note that L2 recurrent connections can be both excitatory and inhibitory, a departure from biology which allows for an easier implementation.

Finally, the recurrent connections in L2 are associated to synaptic delays: for each pair $(i,k)$ of connected L2 neurons, the random delay $\Delta_{ki}$ is drawn uniformly between $0$ and $5$ msec. This substantially increases the diversity of neural responses in the population, particularly at the level of JPSTHs (Figure 3e from the main text)---
this is interesting because our method is specifically designed to analyse generic, heterogeneous population activities.\\

We implemented and simulated the network using Brian, a spiking neural network simulator in Python \citep{Goodman2008}. Our simulation consisted of many successive epochs of 500 msec with all possible successions of the three stimulus values $f$ (as in Figure 1a from the main text). 
Since the input Poisson neurons were always firing close to 30 Hz, there was no strong transient at stimulus onset as is often observed in real sensory neurons. In our case, the change of activity between two successive stimuli was always only differential, and rather weak (see Figure 3c from the main text). 

\begin{table}[tp]
\centering
\begin{tabular}{c c c c c c}

\hline\hline
\hspace*{1mm}\\[-1.5ex]

Subtype & $I^{(s)}$ & $w_\mathrm{min}^{(1,s)}$ & $w_\mathrm{max}^{(1,s)}$ & $w_\mathrm{min}^{(2)}$ & $w_\mathrm{max}^{(2)}$ \\[0.5ex]
\hline \vspace{+0.5mm} \\

Pos. biased ($p$) & 0   &  0  &  2  & -2  &  2 \\
Neg. biased ($n$) & 14  & -3  &  0  & -2  &  2 \\
Unbiased ($u$) & 5     &  0   &  0  & -2  &  2 \\[2ex]

\hline\hline
\end{tabular}
\caption{\label{tab:L2params}Connectivity parameters in the three subtypes of L2 neurons. All values are expressed in millivolts.}
\end{table}

\newpage
\section{Singular value analysis}
\label{sec:suppSVD}

We detail here our mathematical analysis to understand the evolution of SNR and PCV estimates in growing populations of size $\nK$, as a function of the underlying structure of the full population. These results expand the condensed presentation proposed in appendix B of the main text. 

\subsection{Notations}

\subsubsection{Activity across neurons, stimuli and trials}

For simplicity, we consider a timeless version of neural activities, although the whole analysis could be extended to include time as well. In our readout framework, this means that we fix some candidate temporal integration parameters $(w,\tR)$, and consider the resulting neural activities $S_i$, constructed from the temporal integration of each neuron $i$'s spikes\footnote{$S_i$ is noted $\overline{s_i}$ in the main text.}.

Since our main results have been presented in the case of linear tuning to stimuli, we stick to this hypothesis. This implies that all signal/noise properties can be understood by considering only two stimuli (as the difference in response between these two stimuli totally defines the linear tuning of each neuron). We thus note $f=\{0,1\}$ the two possible stimulus values which can be input to the network.

Finally, we may want to consider the possibility of imprecise neural measurements, due to recording from only a finite number of trials (although it is not the main concern of this note). We thus denote $\om\in\Ome$ the set of all possible different realizations of network activity. In theory, $\Ome$ is an infinite set of possible events. However, we will formally assume it to be finite, with (huge) cardinality $\Omega$---so on a given trial, each possible network realization $\om$ has a probability $1/\Omega$ of coming out.

We thus summarize all possible network realizations through the array $S_i^{f\om}$, where $i=1\dots \nN$ denotes all neurons in the population\footnote{$N$ is noted $\Ntot$ in the main text.}, $f=0,1$ denotes stimulus value, and $\om = 1\dots \Omega$ denotes all possible realizations. The notation $f\om$, somewhat abusive, applies the same indexing $\om$ for possible realizations in both stimulus conditions $f=0$ and $f=1$---which can only be done if both stimulus conditions allow the same number $\Omega$ of possible network realizations. However, given the formal nature of ensemble $\Ome$, this notation abuse appears harmless.

As we start doing statistics across neurons and trials, we will need to compute expectancies (i.e., means) and covariance structures across various dimensions. In all cases, we apply the generic notation $\pE_\alpha^A(X_{\alpha,\beta,\dots})$ to denote the empirical mean of quantity $X_{\alpha,\beta,\dots}$ when $\alpha$ is varied over ensemble $A$ ($\beta,\dots$ being any other parameters that are held fixed). When ensemble $A$ is unambiguous, meaning that it includes all possible values for $\alpha$, we will omit it. Finally, second order variances and covariance structures will generically be computed as $\Cov_\alpha^A(X_\alpha,Y_\alpha)=\pE_\alpha^A(X_\alpha Y_\alpha)-\pE_\alpha^A(X_\alpha)\pE_\alpha^A(Y_\alpha)$.

As a first application of these notations, remember that the whole sensitivity analysis derived in the main text deals only with variations: the ``signal'' measures variations of activity with a change in stimulus $f$, while the ``noise'' measures variations of activity across trials $\om$. Thus, the overall mean level of activity for each neuron $i$, that is $\pE_{f\om}(S_i^{f\om})$, plays no role in the analysis: it always disappears from the computations of tuning and noise covariance structure. To clarify further notations, we can thus offset all neural signals and assume that $\pE_{f\om}(S_i^{f\om})=0$, for every neuron $i$ in the population.

\subsubsection{Modes of activity in the neural population}
\label{sec:svd}

The key argument of this note relies on interpreting $S_i^{f\om}$ as a very large $\nN\times(2\Omega)$ matrix, and considering its singular value decomposition (SVD). The (compact) SVD is a standard decomposition which can be applied to any rectangular matrix ${\bf S}$. It writes ${\bf S}=\bU\bLam \bV^\top$, where $\bLam$ is an $M\times M$ diagonal matrix with strictly positive entries $\lambda_m$ (the singular values), $\bU$ is an $\nN\times M$ matrix of orthogonal columns (meaning $\bU^\top \bU=\Id_M$), and $\bV$ is an $\Omega\times M$ matrix of orthogonal columns (meaning $\bV^\top \bV=\Id_M$).

With our current definition of neural activity $S$, the SVD decomposition writes
\begin{equation}
\label{eq:svd}
S_i^{f\om}= \sum_{m=1}^M \lambda_m u_i^m v_m^{f\om},
\end{equation}
where the orthogonality of $\bU$ writes:
\begin{equation}
\label{eq:orthu}
\forall\ (m,n),\; \sum_{i=1}^N\; u_i^m u_i^n = \delta^{mn},
\end{equation}
and the orthogonality of $\bV$ similarly writes $\sum_{f\om}(v_m^{f\om} v_n^{f\om})= \delta_{mn}$. In the case of $\bV$, our above convention that $\pE_{f\om}(S_i^{f\om})=0$ for all neurons $i$ actually imposes that $\pE_{f\om}(v_m^{f\om})=0$ for all modes $m$. We thus reinterpret the orthogonality of $\bV$ as a linear independence between the different random variables $v_m$:
\begin{align}
\forall m,\; & \pE_{f\om}\big(v_m^{f\om})=0, \\
\forall\ (m,n),\; & \Cov_{f\om}\big(v_m^{f\om},v_n^{f\om})= \delta_{mn}. \label{eq:orthv}
\end{align}
Note that we reinterpret the sum over trials ($f,\om$) as an expectancy (thus rescaling $\lambda_m$ by ensemble size $2\Omega$). This allows to emphasize the statistical interpretation of the SVD decomposition in this case.\\

Each triplet $(\lambda_m,\bu^m,v_m)$ defines one particular {\it mode} of activity in the population. We call $\lambda_m$ the {\it power} of the mode, $\bu^m$ (viewed as an $N$-dimensional vector) its {\it distribution vector}, and $v_m$ (viewed as a scalar random variable) its {\it appearance variable}. The appearance variable $v_m$---which takes a different value $v_m^{f\om}$ on every repetition of the experiment--- describes the probability of appearance of each mode $m$ across stimuli and trials. Through \refeq{orthv}, each mode $m$ verifies $\pE_{f\om}\big((v_m^{f\om})^2\big)=1$, meaning that all modes have the same overall ``expected appearance'' across trials. 

Similarly, \refeq{orthu} implies that $\sum_i\big((u^m_i)^2\big)=1$, so $\bu^m$ describes the normalized distribution of the mode across the neural population. Some modes $m$ may correspond to a rather homogeneous distribution of $(u^m_i)^2$ across the population, meaning that the mode is very {\it distributed}, whereas other modes may have power concentrated only over a small subensemble of neurons. These are the modes corresponding to local patterns of activity which only impact a small fraction of the total neural population.

Finally, the power $\lambda_m$ describes the overall impact of mode $m$ on population activity. Indeed, although distribution vectors $\bu^m$ and appearance variables $v_m$ display the same normalization across modes, this does not mean that all modes are equivalent. Instead, only those modes with the largest values $\lambda_m$ will truly impact the population, in the form of measurable changes of activity across neurons and trials. Conversely, modes with small values $\lambda_m$ will scarcely impact population activity, either because they involve only a small fraction of neurons, either because they are distributed but very weak. 

The overall number of modes $M$ is equal to the rank of matrix $S_i^{f\om}$, so it is by construction smaller or equal to the population size $\nN$ (which we assume to be smaller than the huge number $\Omega$ of possible realizations across trials). $M$ defines the typical dimension of the manifold in which all neural activity occurs. In real neural populations, although $\nN$ is itself a very large number, there are reasons to believe that $M$ is sensibly smaller, due to correlated activity between neurons.

\subsubsection{Statistics of activity}

We now reinterpret classical measures of neural activity in the framework defined above. At this point, we need to carefully specify the nature of the ensembles truly available for measures: a finite subset $\Kap$ of neurons from the population, and a finite ensemble $\eE$ of trials (each element of $\eE$ providing one realization for stimulus $f=0$ and one realization for $f=1$).

For every neuron $i\in\Kap$, recorded over trials $\om\in\eE$, we compute the tuning to stimulus as
\begin{equation}
\label{eq:tun-i}
b_i^\eE:=\frac{1}{2}\Big(\pE^\eE_\om(S_i^{1\om})-\pE^\eE_\om(S_i^{0\om})\Big),
\end{equation}
that is, the difference between the experimental mean firing rates in stimulus conditions $f=1$ and $f=0$.
\footnote{Vector $\bb$ from this appendix corresponds to $\sigma_f \bb$ from the main text, where $\sigma_f^2=\langle f^2 \rangle_f-\langle f \rangle_f^2$ gives typical variations of input stimulus.}
Similarly, we compute the noise covariance term between any two neurons $i$ and $j$ as:
\begin{equation}
\label{eq:cov-ij}
C_{ij}^\eE:=\frac{1}{2}\Big(\Cov^\eE_\om(S_i^{0\om},S_j^{0\om})+\Cov^\eE_\om(S_i^{1\om},S_j^{1\om})\Big),
\end{equation}
that is, the stimulus-averaged noise covariance between $i$ and $j$. Finally, we introduce the total covariance matrix $A_{ij}^\eE$ summing up all sources of variance across the population:
\begin{align}
A_{ij}^\eE :&= \Cov_{f\om}^\eE\Big(S_i^{f\om},S_j^{f\om}\Big)\nonumber\\
   &= C_{ij}^\eE + b_i^\eE b_j^\eE. \label{eq:tot-ij}
\end{align}
The last line provides the classic decomposition of the total covariance matrix into noise covariance matrix $\bC^\eE$ and signal covariance matrix $(\bb^\eE)(\bb^\eE)^\top$---which has rank 1 under our assumption of linear tuning to stimulus.

When ensemble $\eE$ is equal to the full space $\Ome$ of possible realizations, the above formulas define the ``true'' measures of covariance, as would be obtained given a sufficient amount of trials. In the sequel, we refer to these true, error-free values, by removing the mention to $\eE$. That is: $b_i$, $C_{ij}$ and $A_{ij}$.\\

The SVD decomposition (\refeq{svd}) is best interpreted as a change of variables reexpressing neural activities $\{S_i\}_{i=1\dots \nN}$ in terms of mode appearance variables $\{v_m\}_{m=1\dots M}$. As a result, we can define the respective equivalents of tuning, noise covariance and total covariance in the space of activity modes. Indeed, although mode appearance variables $v_m$ are never directly observed, they still have some statistics across trials. We thus define:
\begin{align*}
\eta_m^\eE :&=\frac{1}{2}\Big(\pE^\eE_\om(v_m^{1\om})-\pE^\eE_\om(v_m^{0\om})\Big),\\
\Phi_{mn}^\eE :&= \Cov_{f\om}^\eE\Big(v_m^{f\om},v_n^{f\om}\Big),
\end{align*}
which define tuning and total covariance in mode space (noise covariance being implicitly defined as $\bPhi^\eE-(\boeta^\eE)(\boeta^\eE)^\top$). Again, we will denote the true tuning and covariance by removing the mention to $\eE$: true tuning $\boeta$ and true total covariance $\bPhi$. Importantly, the normalization of variables $v_m$ in \refeq{orthv} implies that $\bPhi=\Id_M$.

Mode powers $\lambda_m$ and distribution vectors $\bu^m$ then allow to relate the statistics at the levels of neurons and modes. Injecting the SVD formula (\refeq{svd}) into equations~\ref{eq:tun-i} and \ref{eq:tot-ij} yields 
(in matricial form):
\begin{align}
\bb^\eE &=\bU \bLam \boeta^\eE, \label{eq:b-svd}\\
\bA^\eE &= \bU \bLam \bPhi^\eE \bLam \bU^\top. \label{eq:A-svd}
\end{align}
In particular, when true noiseless measures are considered so that $\bPhi=\Id_M$, we see that $\bU$ and $\bLam$ directly provide the standard (nonzero) eigenvalue decomposition of the total covariance matrix $\bA$, as
$$
\bA = \bU \bLam^2 \bU^\top.
$$

\subsection{SNR and PCV predictions}

We now wish to understand which factors determine the evolution of curve $Z(K)$, the average SNR embedded in neural subensembles $\Kap$ of cardinal $\nK$. We can also study the evolution of percept covariance (PCV) signals, in the same framework.

In the main text, we compute SNR and PCV for ensemble $\Kap$ through Fisher's linear discriminant (eq. 13-16). 
 One sees easily that these definitions, involving tuning $\bb$ and noise covariance matric $\bC$, are equivalently expressed in terms of tuning $\bb$ and {\it total} covariance matrix $\bA$:
\begin{align}
\ba_\Kap&= (\bb_\Kap^\top \bA_\Kap^{-1} \bb_\Kap^{~})^{-1}\;\bA_\Kap^{-1} \bb_\Kap^{~},\label{eq:a-kap}\\
Y(\Kap) &= \bb_\Kap^\top \bA_\Kap^{-1} \bb_\Kap^{~}.\label{eq:Y-kap}
\end{align}
We call $Y$ the signal-to-total ratio (STR), which relates directly to SNR $Z$ by the formula $Y=Z/(1+Z)$. $Y$ always takes values between $0$ ($Z=0$) and $1$ ($Z=\infty$), it thus avoids singularities which may occur in the direct $Z$ formulation. If matrix $\bA_\Kap$ is rank-deficient, we consider its (Moore-Penrose) pseudoinverse without loss of generality (see further down).



\subsubsection{Total STR in the population}

The SVD decomposition (\refeq{svd}) reexpresses neural activity in the space of modes $m=1\dots M$. When the full neural population is considered, the full matrix $\bA$ and vector $\bb$ are involved in \refeq{Y-kap}. Using the SVD formulations (\refeq{b-svd}-\ref{eq:A-svd}) we thus find:
\begin{align}
Y(\infty)&=\bb^\top \bA^{-1} \bb\nonumber\\
&=\boeta^\top \bLam \bU^\top (\bU \bLam ^2 \bU^\top)^{-1} \bU \bLam \boeta\nonumber\\
&=\|\boeta\|^2= \sum_{m=1}^M \eta_m^2.\label{eq:Y-infty}
\end{align}
Thus, each mode contributes to total sensitivity by the strength of its intrinsic sensitivity $\eta_m$.

This computation can also be derived assuming a finite number of experimental trials $\eE$. In this case however, we must introduce the {\it experimental sensitivity} $\zeta^\eE_m$ of each mode, defined as
\begin{align}
\label{eq:bzeta}
\bzeta^\eE := (\bPhi^\eE)^{-\frac{1}{2}} \boeta^\eE,
\end{align}
where $(\bPhi^\eE)^{-\frac{1}{2}}$ is the unique (Moore Penrose) pseudo-inverse of the symmetric, non-negative square root matrix of $\bPhi^\eE$. Actually, any other choice of matrix square root could also be used, because by construction $\bPhi^\eE \succeq (\boeta^\eE)(\boeta^\eE)^\top$, in the sense of symmetric positive matrices. This insures that $\boeta^\eE$ is orthogonal to $\mathrm{Ker}(\bPhi^\eE)$, and thus the unicity of $\bzeta^\eE$ as defined in \refeq{bzeta}.

The computation of $Y(\infty,\eE)$ then goes along the same lines as previously:
\begin{align}
Y(\infty,\eE)&=(\bb^\eE)^\top (\bA^\eE)^{-1} \bb^\eE \nonumber\\
&=(\boeta^\eE)^\top \bLam \bU^\top (\bU \bLam \bPhi^\eE \bLam \bU^\top)^{-1} \bU \bLam \boeta^\eE\nonumber\\
&=\|\bzeta^\eE\|^2.\label{eq:Y-infty-E}
\end{align}
Generally, one expects $Y(\infty,\eE)>Y(\infty)$, because the estimated $\bPhi^\eE$ is flatter than its true value of $\bPhi=\Id_M$, with eigenvalues closer to 0.
This is a classic result when estimating SNR (or STR) from an insufficient number of trials, a typical example of overfitting. As mentionned in the main text, there is no miracle cure to this problem, which should be addressed through appropriate methods of regularization and cross-validation\citep{Hastie2009}.

\subsubsection{STR for finite neural ensembles}

We now turn to the sensitivity embedded in finite subensembles $\Kap$ from the population. The definitions of $\bA_\Kap$ and $\bb_\Kap$ used in \refeq{Y-kap} amount to a projection from the full neural space $\mathbb{R}^N$ to subensemble $\Kap$:
\begin{align*}
\bb_\Kap &=\bP_\Kap \bb,\\
\bA_\Kap &=\bP_\Kap \bA \bP_\Kap^\top,
\end{align*}
where $\bP_\Kap$ is the $\nK\times\nN$ orthogonal projector on recorded neurons $\Kap$. Through the SVD decomposition in \refeq{b-svd}-\ref{eq:A-svd}, we reexpress these quantities as:
\begin{align}
\bb_\Kap 
&= \bD_\Kap^\top \boeta\label{eq:bK-svd}\\
\bA_\Kap 
&= \bD_\Kap^\top \bD_\Kap\label{eq:AK-svd}, 
\end{align}
where
\begin{align}
\bD_\Kap := \bLam \bU^\top \bP_\Kap^\top,\label{eq:D-kap}
\end{align}
is our so-called {\it data matrix}, an $M \times \nK$ matrix with elements $d_i^m:=\lambda_m u^m_i$. It represents the experimental data from neurons $\Kap$, expressed in mode space.

To compute the resulting sensitivity predicted by \refeq{Y-kap}, we note that through \refeq{AK-svd}, matrix $\bA_\Kap$ has the same eigenvalues as its dual Gram matrix $\bD_\Kap \bD_\Kap^\top$, an $M \times M$ matrix with rank $d:=\mathrm{min}(\nK,M)$---generally equal to $\nK$. We introduce the (compact) SVD decomposition of this matrix:
$$
\bD \bD^\top= \bX \bT^2 \bX^\top,
$$
where $\bT^2>0$ is a $d \times d$ diagonal matrix, and $\bX$ is an $M \times d$ matrix of orthogonal columns (for clarity we remove the unambiguous references to ensemble $\Kap$). It is shown easily that this decomposition also provides the SVD for $\bA_\Kap$, in the form:
$$
\bA_\Kap = (\bD^\top\bX\bT^{-1}) \bT^2 (\bD^\top\bX\bT^{-1})^\top,
$$
where $(\bD^\top\bX\bT^{-1})$ is a $\nK \times d$ matrix of orthogonal columns, as required in the SVD decomposition. Thus, the (pseudo-)inverse of $\bA_\Kap$ writes:
$$
\bA_\Kap^{-1} = (\bD^\top\bX\bT^{-1}) \bT^{-2} (\bD^\top\bX\bT^{-1})^\top.
$$
This allows to finally compute the experimental STR, from \refeq{bK-svd}-\ref{eq:AK-svd}:
\begin{align*}
Y(\Kap) &= \bb_\Kap^\top\bA_\Kap^{-1}\bb_\Kap\\
&= \boeta^\top \bD (\bD^\top\bX\bT^{-1}) \bT^{-2} (\bD^\top\bX\bT^{-1})^\top \bD^\top \boeta\\
&= \boeta^\top \bX\bT^{2}\bX^\top \bX \bT^{-4} \bX^\top \bX \bT^{2} \bX^\top \boeta\\
&= \boeta^\top (\bX \bX^\top) \boeta,
\end{align*}
making use of the fact that $\bX^\top \bX=\Id_d$. Intriguingly matrix $\bT$, which describes the eigenvalues of $\bA_\Kap$, disappears from the final equation. Only matrix $\bX$, corresponding to the {\it eigenvectors} of $\bD\bD^\top$, remains in the equations. We note
$
\bDelta_\Kap := \bX_\Kap \bX_\Kap^\top,
$
which is nothing but the $M \times M$ orthogonal projector on $\mathrm{Im}\big(\bD_\Kap)$. This leads to the final expression:
\begin{equation}
\label{eq:Y-kap-final}
Y(\Kap) = \boeta^\top \bDelta_\Kap \boeta.
\end{equation}

Neuron ensemble $\Kap$ only appears through $\bDelta_\Kap$. 
In particular, as soon as $\nK$ is larger than the number of modes $M$, necessarily $\bDelta_\Kap=\Id_M$, and $Y(\Kap)=Y(\infty)$: all modes are available experimentally, and sensitivity estimates saturate to their maximum value, independently of ensemble $\Kap$.\\

The whole analysis can be performed similarly assuming a finite number of measurement trials $\eE$. The only difference is a modification in data matrix $\bD$, to take into account the biases in mode space induced by an insufficient number of trials:
$\bD_\Kap^\eE := (\bPhi^\eE)^{\frac{1}{2}} \bLam \bU^\top \bP_\Kap^\top,
$ 
using the same square root of $\bPhi^\eE$ as in \refeq{bzeta}. Similar computations lead to the final result:
\begin{equation}
Y(\Kap,\eE) = (\bzeta^\eE)^\top \bDelta_\Kap^\eE \bzeta^\eE,
\end{equation}
which depends on experimental mode sensitivities (\refeq{bzeta}) and on $\bDelta_\Kap^\eE$, the orthogonal projector on $\mathrm{Im}\big(\bD_\Kap^\eE)$, of dimension $d=\mathrm{min}(\nK,M,E)$.

\subsubsection{Percept covariance for finite readout ensembles}
\label{sec:appPCV}

Similarly to the approach above, we can express PCV signals in mode space. Since we do not model time, we only have access to the temporal average $\overline{\pi_i}:=\int_{u>0} \pi_i(\tR-u)h_w(u)du$, where $\pi_i(t)$ is the full PCV curve from the main text. From eq. 9 of the main text, 
it falls easily that $\overline{\bpi}=\bC \ba$.
Using the optimal $\ba$ for readout ensemble $\Kap$ (\refeq{a-kap}, with $\ba=\bP_\Kap^\top \ba_\Kap^{~}$ since $\ba$ has support on $\Kap$), we thus predict:
$$
\overline{\bpi}(\Kap) = Y(\Kap)^{-1}\;\bC \bP_\Kap^\top \bA_\Kap^{-1} \bb_\Kap^{~},
$$
which provides the value of $\overline{\pi_i}$ for every neuron $i$ in the population (not only in ensemble $\Kap$). Making use of the same SVD decompositions as above, and of relationship $\bC=\bA-\bb \bb^\top$, we finally find:
\begin{equation}
\label{eq:pi-kap-final}
\overline{\bpi}(\Kap)+\bb=Y(\Kap)^{-1}\; \bU \bLam \bDelta_\Kap \boeta,
\end{equation}
which expresses $\overline{\bpi}(\Kap)$ as a linear combination of mode distribution vectors $\bu^m$. As $\Kap$ tends to the full population, $\bDelta_\Kap$ tends to $\Id_M$ and we get $\overline{\bpi}(\infty)=Y^{-1}\bb-\bb=Z^{-1}\bb$, the prediction for choice signals in case of optimal readout \citep{Haefner2013}.

In turn, the population average for PCV is $\Wbar(\Kap):=\pE_i(b_i \overline{\pi_i}(\Kap))$\footnote{which corresponds to the temporal integral $\int_{u>0} W(\tR-u|\Kap)h_w(u)du$ for the PCV curve $W(t|\Kap)$ defined in the main text (eq. 16).}. 
Using \refeq{pi-kap-final}, and the general fact that $\pE_i(x_iy_i)=N^{-1}\mathbf{x}^\top \mathbf{y}$, we obtain
\begin{align}
\Wbar(\Kap)+\pE_i(b_i^2) &=  N^{-1}\bb^\top\big(\overline{\bpi}(\Kap)+\bb\big)\nonumber\\
&= (NY(\Kap))^{-1} \boeta^\top \bLam^2 \bDelta_\Kap \boeta, \label{eq:W-kap-final}
\end{align}
because $\bb=\bU \bLam \boeta$ (\refeq{b-svd}) and $\bU^\top \bU=\Id$.
This reveals the interest of multiplying $\overline{\pi_i}$ by the corresponding tuning $b_i$ (see discussion in main text): it allows to get rid of the unknown distribution vectors $\bU$, and instead produce a quantity $W$ which is directly related to the underlying modes' powers $\bLam$ and sensitivities $\boeta$. 

As it appears in \refeq{W-kap-final}, we note $B^2:=\pE_i(b_i^2)$ the average square tuning in the population. With similar arguments as above, one shows that \begin{equation}
\label{eq:B2}
B^2=N^{-1}\boeta^\top \bLam^2 \boeta=N^{-1}\sum_{m=1}^M\eta_m^2 \lambda_m^2.
\end{equation}


\subsubsection{Behavior with $K$}

We are now better armed to understand how sensitivity and PCV predictions vary as a function of the readout ensemble $\Kap$. We are mostly interested in averages of these quantities over randomly chosen ensembles $\Kap$ of size $K$~; we thus use the generic notation $\pE_K(x):=\pE(x(\Kap)|\mathrm{Card}(\Kap)=K)$. From \refeq{Y-kap-final} we find: $\pE_K Y=\boeta^\top (\pE_K \bDelta) \boeta$.

To understand the properties of the $(M \times M)$ matrix $\pE_K \bDelta$, we view the $(M \times K)$ data matrix $\bD_\Kap$ (\refeq{D-kap}) as a collection of $K$ random vectors $\bd_i$ in mode space, viewing neuron identities $i$ as the random variable. Thus, $\bDelta_\Kap$ is the orthogonal projector on the linear span of the $K$ sample vectors $\{\bd_i\}_{i \in \Kap}$. As a projector, its trace is equal to its rank, so we have $\mathrm{Tr}(\pE_K \bDelta)=K$. Furthermore, since $K+1$ samples span on average more space than $K$ samples, we are insured that $\pE_{K+1} \bDelta \succeq \pE_K \bDelta$, in the sense of positive definite matrices.
Finally, intuition and numerical simulations suggest that $\pE_K \bDelta$ is almost diagonal. Indeed, as the various modes are linearly independent (\refeq{orthu}), there is no linear interplay between the different dimensions of $\bd$ across samples $i$: $\pE_i(d_i^m d_i^n)=N^{-1}\lambda_m^2 \delta^{mn}$, or equivalently 
$$\pE_K(\bX_\Kap \bT_\Kap^2 \bX_\Kap^\top)=\pE_K(\bD_\Kap \bD_\Kap^\top)=KN^{-1} \bLam^2.$$
Assuming a form of independence between $\bX$ and $\bT$, it is reasonable to suppose that $\pE_K(\bX_\Kap \bX_\Kap^\top)=\pE_K \bDelta$ is close to diagonal as well\footnote{A rigorous proof might be accessible assuming a normal distribution for random vector $\bd$. In the general case, small deviations from diagonality can probably occur.}.

Assuming that $\pE_K \bDelta$ is diagonal, we note its diagonal terms $\{\epsilon^m_K\}_{m=1\dots M}$ and consider the resulting approximations of sensitivity (\refeq{Y-kap-final}) and mean PCV (\refeq{W-kap-final}):
\begin{align}
\pE_K Y &\simeq \sum_{m=1}^M \epsilon^m_K \eta_m^2,\label{eq:EKY}\\
\pE_K \Big(Y(\Wbar+B^2)\Big) &\simeq N^{-1}\sum_{m=1}^M \epsilon^m_K \lambda_m^2 \eta_m^2.\label{eq:EKWY}
\end{align}
The properties of $\pE_K \bDelta$ imply that $\sum_m \epsilon^m_K=K$ (trace property), and $\epsilon^m_{K+1}\geq\epsilon^m_K$ (growth property).
As $K$ augments, $\{\epsilon^m_K\}$ progressively ``fills-in'' the space of modes, starting from the modes with larger power $\lambda_m$. Indeed, the larger $\lambda_m$, the more often mode $m$ appears in samples $\{\bd_i\}$. As a useful image, we may think of the (very) rough approximation $\epsilon^m_K \simeq \mathbb{1}_{m\leq K}$: only the $K$ first modes are revealed by a sample of $K$ neurons. Naturally this is only a gross approximation, as can be seen easily by considering a single sample $i$ ($K=1$). From intuition and simulation, the true shape of $\{\epsilon^m_K\}$ (at fixed $K$) is a ``smoothed'' version of $\mathbb{1}_{m\leq K}$, and the degree of smoothing depends on the power law governing the spectrum $\{\lambda_m\}$.

With this image in mind, \refeq{EKY} shows that the growth of sensitivity with $K$ is linked to the progressive summation of mode sensitivities $\eta_m^2$, starting from modes with highest power $\lambda_m$:
$$
\pE_K Y\;\; \underset{K}{\nearrow} \;\; Y(\infty),
$$
with a saturation as soon as all nonzero mode sensitivities $\eta_m$ are revealed.
Conversely, for PCV signals, we can make the rough assumption that $\pE_K(\Wbar Y)\simeq\pE_K(\Wbar)\pE_K(Y)$, in which case \refeq{EKWY} rewrites
$$
\pE_K \Wbar + B^2 \simeq N^{-1} \frac{\sum_{m=1}^M \epsilon^m_K \lambda_m^2 \eta_m^2}{\sum_{m=1}^M \epsilon^m_K \eta_m^2}
:= \Big\langle \frac{\lambda^2_m}{N} \Big\rangle_{m,K},
$$
where each mode $m$ contributes with a weight $\epsilon^m_K \eta_m^2$, and $\pE_K Y=\sum_m \epsilon^m_K \eta_m^2$ provides the normalization factor. Thus, $\langle \lambda^2_m \rangle_{m,K}$ reflects the average power of modes with the higher sensitivity, that are already revealed with $K$ neurons. As $K$ grows, $\{\epsilon^m_K\}$ progressively ``fills-in'' modes in the order of decreasing $\lambda_m$. Thus we expect $\langle \lambda^2_m \rangle_{m,K}$ to decrease with $K$. Finally, as soon as $K\geq M$, we have $\{\epsilon^m_K\}=\{1\}$, and
$$
\Big\langle \frac{\lambda^2}{N} \Big\rangle_{m,\infty}=N^{-1}\frac{\sum_{m=1}^M \lambda_m^2 \eta_m^2}{\sum_{m=1}^M \eta_m^2}=\frac{B^2}{Y(\infty)},
$$
reckognizing the expressions for $B^2$ (\refeq{B2}) and $Y(\infty)$ (\refeq{Y-infty}). Since $Y^{-1}-1=Z^{-1}$, the predicted evolution of mean PCV signal with $K$ follows:
$$
\pE_K \Wbar\;\; \underset{K}{\searrow} \;\; \frac{B^2}{Z(\infty)} > 0.
$$
$\Wbar$ is predicted to be positive, to decrease with increasing size $K$, and to saturate at its minimum value once all significant mode sensitivities $\eta_m$ have been revealed---which is also the moment when sensitivity $Y$ saturates at its maximum value (\refeq{EKY}), and corresponds to an optimal readout from the full population. The implications of these results in terms of extrapolation to large $K$ are discussed in the main text.

\end{document}